\documentclass[a4paper,fleqn,usenatbib]{mnras}

\usepackage{natbib} %,widetext}
\usepackage{amsmath,amssymb}
\usepackage{epsfig,graphicx}
\usepackage{epstopdf}
\usepackage{subcaption}
\usepackage{float}
\usepackage{color}
\usepackage{lastpage}
\DeclareGraphicsExtensions{.pdf,.png,.jpg,.eps}
\newcommand{\wgg}{\ensuremath{w_{gg}}}

\newcommand{\xigg}{\ensuremath{\xi_{gg}}}

\newcommand{\mpch}{\ensuremath{h^{-1}\text{Mpc}}}
\newcommand{\mpc}{Mpc}

\def\reff@jnl#1{{\rm#1\/}}

\def\aj{\reff@jnl{AJ}}                  % Astronomical Journal
\def\araa{\reff@jnl{ARA\&A}}            % Annual Review of Astron and Astrophys
\def\apj{\reff@jnl{ApJ}}                        % Astrophysical Journal
\def\apjl{\reff@jnl{ApJ}}               % Astrophysical Journal, Letters
\def\apjs{\reff@jnl{ApJS}}              % Astrophysical Journal, Supplement
\def\apss{\reff@jnl{Ap\&SS}}            % Astrophysics and Space Science
\def\aap{\reff@jnl{A\&A}}               % Astronomy and Astrophysics
\def\aapr{\reff@jnl{A\&A~Rev.}}         % Astronomy and Astrophysics Reviews
\def\aaps{\reff@jnl{A\&AS}}             % Astronomy and Astrophysics, Supplement
\def\baas{\reff@jnl{BAAS}}              % Bulletin of the AAS
\def\jrasc{\reff@jnl{JRASC}}            % Journal of the RAS of Canada
\def\memras{\reff@jnl{MmRAS}}           % Memoirs of the RAS
\def\mnras{\reff@jnl{MNRAS}}            % Monthly Notices of the RAS
\def\physrep{\reff@jnl{Phys.Rep.}}
\def\pra{\reff@jnl{Phys.Rev.A}}         % Physical Review A: General Physics
\def\prb{\reff@jnl{Phys.Rev.B}}         % Physical Review B: Solid State
\def\prc{\reff@jnl{Phys.Rev.C}}         % Physical Review C
\def\prd{\reff@jnl{Phys.Rev.D}}         % Physical Review D
\def\prl{\reff@jnl{Phys.Rev.Lett}}      % Physical Review Letters
\def\pasp{\reff@jnl{PASP}}              % Publications of the ASP
\def\pasj{\reff@jnl{PASJ}}              % Publications of the ASJ
\def\skytel{\reff@jnl{S\&T}}            % Sky and Telescope
\def\solphys{\reff@jnl{Solar~Phys.}}    % Solar Physics
\def\sovast{\reff@jnl{Soviet~Ast.}}     % Soviet Astronomy
\def\ssr{\reff@jnl{Space~Sci.Rev.}}     % Space Science Reviews
\def\nat{\reff@jnl{Nature}}             % Nature

\newcommand{\Msun}{M_{\odot}}

\newcommand{\beq}{\begin{equation}}
\newcommand{\eeq}{\end{equation}}
\newcommand{\beqa}{\begin{eqnarray}}
\newcommand{\eeqa}{\end{eqnarray}}

\graphicspath{{figures/}}%make sure to change eps to pdf before arxiv submission

\definecolor{referee_C}{cmyk}{1,0,0,0}

\newcommand{\referee}[1]{{{ #1}}}

\newcommand{\lcdm}{{$\Lambda$CDM}}

\newcommand{\DS}{\ensuremath{\Delta\Sigma}}

\newcommand{\hdelta}{\ensuremath{\widehat{\delta}}}
\newcommand{\hxi}{\ensuremath{\widehat{\xi}}}
\newcommand{\mean}[1]{\ensuremath{\left\langle #1 \right\rangle}}

\renewcommand{\vec}[1]{\mathbf{#1}}

\title[]{Galaxy-galaxy lensing estimators and their covariance properties}

\author[]{
Sukhdeep Singh$^1$\thanks{\tt sukhdeep@cmu.edu},
Rachel Mandelbaum$^{1}$,%\newauthor
Uro\v{s} Seljak$^{2,3}$,
An\v{z}e Slosar$^4$,\newauthor
Jose Vazquez Gonzalez$^4$\\
$^1$McWilliams Center for Cosmology, Department of Physics, Carnegie Mellon University, Pittsburgh,
PA 15213, USA\\
$^2$Department of Physics, University of California at Berkeley, Berkeley, CA 94705, USA\\
$^3$Lawrence Berkeley National Laboratory, Berkeley, CA 94720, USA\\
$^4$Physics Department, Brookhaven National Laboratory, Upton, NY
11973, USA\\
}

\date{\today}
\date{Accepted XXX. Received YYY; in original form ZZZ}
\pubyear{2016}
\begin{document}
\label{firstpage}
\pagerange{\pageref{firstpage}--\pageref{LastPage}}
\maketitle

\begin{abstract}
We study the covariance properties of real space correlation
function estimators -- primarily galaxy-shear correlations, or galaxy-galaxy lensing --  using SDSS data for both shear
catalogs and lenses (specifically the BOSS LOWZ sample). 
Using mock catalogs of lenses and sources, we 
disentangle the various contributions to the covariance
matrix and compare them with a simple analytical model. We show that not
subtracting the lensing measurement around random points from the
measurement around the lens sample is equivalent to performing the
measurement using the lens density field instead of the lens over-density field. 
\referee{ While the measurement using the lens density field is unbiased (in the absence of systematics), 
 its error is significantly larger due to an additional term in the
covariance. }
Therefore,
this subtraction should be performed regardless of its beneficial effects
on systematics. Comparing the error estimates from data and mocks for estimators that involve the
over-density, we find that the 
errors are dominated by the shape noise and lens clustering, 
that empirically estimated covariances 
(jackknife and standard deviation across mocks) are consistent with theoretical estimates, 
and that both the connected parts of the 4-point function and the super-sample covariance can be
neglected for the current levels of noise. While the trade-off between different terms in
the covariance depends on the survey configuration (area, source number
density), the diagnostics that we use in this work should be useful for
future works to test their empirically-determined covariances.
\end{abstract}

\begin{keywords}
  galaxies: evolution\ --- cosmology: observations
  --- large-scale structure of Universe\ --- gravitational
  lensing: weak
\end{keywords}

\section{Introduction}\label{sec:intro}
	Galaxy-galaxy lensing, the measurement of the tangential shape distortion (``shear'') due to
    gravitational lensing by galaxies, has emerged as an important cosmological probe 
	to study the dark matter distribution around 
	galaxies and the growth of large scale structure \citep{Bartelmann2001,Weinberg2013}. Since lensing is sensitive 
	to all matter, 
	galaxy-galaxy lensing provides a unique way to map the matter distribution around  galaxies (or
    galaxy clusters) and has been measured to 
	good precision by many surveys
    \citep[e.g.,][]{Fischer2000,Hoekstra2004,Sheldon2004,Mandelbaum2006,Heymans2006,Uitert2012,Leauthaud2012,Velander2014,Viola2015,Hudson2015,Clampitt2016,Buddendiek2016}. Several studies 
	have used galaxy-galaxy lensing to study the
	halo mass of galaxies and understand the galaxy-halo connection 
	\citep[e.g.,][]{Hoekstra2004,Mandelbaum2006,Heymans2006,Tinker2012,Leauthaud2012,Uitert2012,Gillis2013,Velander2014,Sifon2015,Hudson2015,Uitert2016}. 
	In combination with galaxy clustering, galaxy-galaxy lensing can also be 
	used to recover the underlying matter correlation 
	function, which can then be used to constrain cosmology 
	\citep{Seljak2005,Baldauf2010,Mandelbaum2013,More2015,Kwan2016,Buddendiek2016} and to test the theory of 
	gravity \citep{Zhang2007,Reyes2010,Blake2016}.
	
	With the increasing precision of cosmological surveys, there has been an increasing focus on estimating the 
	covariances of the measurements more accurately as well, since the uncertainties in covariance matrices can lead to 
	incorrect estimation of uncertainties in cosmological parameters
	\citep[e.g.,][]{Hartlap2007,Dodelson2013,Taylor2013,Mohammed2016}. 
	Estimating the non-Gaussian or connected part of the covariance
    matrix, which has two 
	contributions, is especially challenging. The first contribution is due to mode couplings between small-scale (or in-survey) modes from 
	the non-linear evolution of structure \citep{Scoccimarro1999,Hu2001,Cooray2001,Mohammed2016}. 
	The second term is  the super-sample variance contribution from the 
	couplings of modes within the survey to the modes corresponding to length scales that are
    larger than the survey size 
    \citep[eg. ][]{Hu2003,Hamilton2006,Takada2013,Takada2014,Schaan2014,Li2014}. Current prescriptions for 
    estimating covariances include using numerical simulations \citep[e.g.,][]{Dodelson2013,Manera2013,Li2014}, 
    using a physically-motivated halo model \citep[e.g.,][]{Takada2013,Schaan2014} and using
    analytical estimates based on 
    perturbation theory \citep[e.g.,][]{Mohammed2016}.
	
	While many studies have explored this issue of covariance matrices for two-point functions in general,
    galaxy-galaxy lensing covariances have been relatively less well-studied. When addressing this question, one 
    must also address the question of which estimator is used for the measurement. 
    Several estimators for the galaxy-galaxy lensing signal can be found in
    the literature.  One estimator uses  the average tangential shear of
    background galaxies with respect to the lens galaxies.  Another estimator also includes the
    subtraction of tangential shear around random points, which has an expectation value of zero in
    the absence of systematics and which can be used to remove the impact of coherent additive shear
    systematics.  Subtraction of the lensing shear around random points is often argued
    to be beneficial primarily due to the way it removes these additive systematic errors  \citep[e.g.,][]{Mandelbaum2005,Mandelbaum2013}.  However, another motivation
    for the latter estimator can be found in the work on optimal estimators of galaxy clustering:
    \cite{Landy1993} illustrated that the estimators that are constructed using mean-zero quantities (over-density), 
    while having an expectation value that is the same as the simple estimator, have better covariance
    properties\footnote{Sometimes the reverse claim is made in the literature -- i.e., that the
      estimator with the signal around random points subtracted has increased variance.  This claim
      is typically made in cases
      where not enough random points are used, in which case there is indeed some added variance.  Our
      argument that the covariance properties of this estimator are superior is true in the limit of infinite random
      points: they are nothing other than a Monte Carlo method to determine the survey volume and hence the mean 
      density. We explore this issue in more realistic cases in this work.}.
      For example, in the case of galaxy surveys, random points ($R$) that follow the area coverage
      of the lenses are typically used to estimate the mean of the galaxy field 
      ($D$) 
      in the presence of complicated survey masks.  They are then used to convert the galaxy field into
      the normalized over-density (mean-zero)
      field $(D-R)/R$, the auto-correlation of which is the standard Landy-Szalay estimator for
      galaxy clustering with improved covariance properties.   Likewise, for galaxy-galaxy lensing,
      estimating the mean tangential shear around lens galaxies corresponds to correlating the
      galaxy density field (nonzero mean) with the shear field, while subtraction of the mean
      tangential shear around random points results in correlating the
      mean-zero galaxy over-density with the shear.
   
	In general, most galaxy-galaxy lensing studies either compute the covariance matrices analytically assuming shape 
	noise and measurement noise only \citep[see for example, ][]{Viola2015}, or
	use the jackknife method, which has the advantage that it includes all observational effects, though it is
	noisier and also limits the scales which can be used in the analysis \referee{\citep[e.g., see][for comparison of 
	theoretical and jackknife covariance]{Hildebrandt2016,Blake2016}}. 
	It is also not clear how well the jackknife 
	method can capture the super-sample covariance, though since galaxy-galaxy lensing is dominated by shape noise in 
	current generation surveys, super-sample covariance is expected to be subdominant.
	Recently \cite{Shirasaki2016} \referee{\citep[see also][]{Blake2016}} did a detailed 
	study of the galaxy-galaxy lensing covariance matrix using realistic $N$-body and ray tracing simulations. 
	\referee{In version 1 of their paper, } they found 
	that once the scales are of similar order as the jackknife division size, the jackknife method overestimates the 
	errors 
	compared to errors obtained from the standard deviation across different simulation realizations, even in the 
	presence of shape noise.
	This overestimation in jackknife errors was interpreted as increased contribution from
    super-sample covariance, since the jackknife method has effectively divided the survey into
    several small survey  
	realizations, and the super-sample covariance grows with the square of the mass variance within the survey volume
	\citep{Takada2013}, which can scale differently from the usual inverse-volume scaling of the covariance terms. 
	The analysis by \cite{Shirasaki2016} applies to the galaxy-galaxy lensing estimators without subtracting the 
	measurement around the 
	randoms lens sample, as in e.g.\ \citet{Leauthaud2012,Viola2015,Hudson2015,Uitert2016,Blake2016} 
	\referee{(\cite{Blake2016} subtracted the measurement around random points from the signal, but
      this was not done for the 
	covariance estimation)}.
	Other galaxy-galaxy lensing studies subtract out the signal around the randoms
	\citep[e.g.,][]{Sheldon2004,Mandelbaum2005,Mandelbaum2006,Clampitt2016,Kwan2016}. 
	\referee{In an updated version of their paper, \cite{Shirasaki2016} show that after subtracting out the measurement around 
	randoms, the covariance decreases and the covariance from the jackknife method is consistent
    with the covariance obtained
	using different mock realizations (for scales smaller than the size of the jackknife regions).}

    In this work, we
    explore the covariance properties of these two galaxy-galaxy lensing estimators both in the
    presence and the absence of systematic errors.  We show that there is a theoretical reason to believe that
    the estimator with the mean shear around random points subtracted should have more optimal
    covariance properties, and we explore the impact of this difference in practice for one particular survey.
	Aside from the issue of removing systematics, we
    demonstrate the correlated noise term between measurements around galaxies and
    randoms, which results in more optimal variance properties after subtracting the shear around
    random points.  We also study the differences in the covariance matrices obtained from the jackknife method and 
	standard deviations across several mock realizations, similar to \cite{Shirasaki2016}.
    We demonstrate several methods of empirically estimating specific covariance
    contributions, and interpret the results of those methods in terms of which galaxy-galaxy
    lensing covariance terms
    they include.
	
	This work is organized as follows. In Sec.~\ref{sec:formalism} we briefly review the theoretical formalism and 
	estimators, and in Sec.~\ref{sec:data} we present the data used. Results are presented in
    Sec.~\ref{sec:results}, and we conclude in \ref{sec:conclusion}. In appendix~\ref{appendix:covariance} we derive the
    expressions for covariance when cross-correlating non-zero mean quantities and in appendix~\ref{appendix:clustering} 
    we present comparisons of different estimators in the case of galaxy clustering measurements.
    
    Throughout this work we use the Planck 2015 cosmological parameters 
	\citep{Planck2015cosmo} with $\Omega_m=0.309$, $n_s=0.967$, $A_s=2.142\times10^{-9}$, $\sigma_8=0.82$. 
	Theory predictions are computed using the linear theory $+$ halofit 
	\citep{Smith2003,Takahashi2012} power spectrum generated with the {\sc CAMB} \citep{Lewis2002} software. \referee{We 
	use $h=1$ when computing distances and hence our $\Delta\Sigma$ measurements are in units of $h\Msun/\text{pc}^2$.}

\section{Formalism and Methodology}\label{sec:formalism}
	\subsection{Galaxy lensing}
		Here we briefly review the formalism of galaxy-galaxy lensing. For a general review of gravitational lensing we 
		refer 
		the reader to \cite{Bartelmann2001,Weinberg2013,Kilbinger2015}.
		
		In galaxy-galaxy lensing, we measure the projected surface mass density
        $\Sigma$ around the lens 
		galaxies. In the case of a spherically symmetric lens, we can write the convergence and shear as
		\begin{align}
			&\kappa(r_p)=\frac{\Sigma(r_p)}{\Sigma_{c}}\\
			&\gamma_t(r_p)=\frac{\bar{\Sigma}(<r_p)-\Sigma(r_p)}{\Sigma_{c}}.
		\end{align}
		$\bar{\Sigma}(<r_p)$ is the mean surface mass density within the transverse separation
        $r_p$, and the critical surface density is
		defined as
		\begin{equation}\label{eqn:sigma_crit}
    		\Sigma_c=\frac{c^2}{4\pi G}\frac{f_k(\chi_s)}{(1+z_l)f_k(\chi_l) f_k(\chi_{s}-\chi_l)},
	    \end{equation}
where	    $f_k(\chi)$ is the transverse comoving distance ($f_k(\chi)=\chi$ in a flat universe). $1+z_l$ converts the 
	    $c^2/G$ factor to comoving space. 
	    
		We can write $\Sigma$ in terms of the projected galaxy-matter correlation function as
		\begin{equation}
    		\Sigma(r_p)=\bar\rho_m\int \mathrm{d}\Pi \, \xi_{gm}(r_p,\Pi)=\bar\rho_m w_{gm}(r_p),
	    \end{equation}
	    where $\Pi$ denotes the line-of-sight separation from the halo center, and we have ignored the effects of lensing
	    window function, which depends on $\Pi$.  Nominally the definition for $\Sigma$ should
        include a factor of $1+\xi_{gm}$
        within the integral (rather than just $\xi_{gm}$), but the constant term does not contribute to $\gamma_t$ because it 
        gets
        removed by subtraction of the $\bar{\Sigma}(<r_p)$ term.
		In the linear bias regime, the galaxy-matter projected correlation function can be derived from the matter power 
		spectrum as
		\begin{align}
			w_{gm}(r_p)=b_g r_{cc}\int \mathrm{d}z W(z)\int \frac{\mathrm{d}^2\vec k}{(2\pi)^2}& P_{\delta
			\delta}(\vec{k},z)e^{i(\vec{r}_p\cdot\vec{k})},\label{eqn:xi}
	    \end{align}
		 where $b_g$ is the galaxy bias and $r_{cc}$ is the galaxy-matter cross correlation
         coefficient, both of which are assumed to 
		 be
		 independent of redshift in this equation. \referee{$P_{\delta\delta}(k,z)$ is the matter power spectrum 
		 (linear$+$halofit) at redshift $z$.} 
		 To lowest order, lensing measurements are not affected by redshift 
		 space distortions,
		 and hence we do not include any corrections for them. The weight function $W(z)$ depends on the redshift
		 distribution of the source galaxies and on the weights used in the estimators
		 when measuring the signal (see Sec.~\ref{ssec:estimators}). We explicitly include these weights when
		 computing the effective redshift $z_\text{eff}$ for the theory calculations.

	\subsection{Estimator}\label{ssec:estimators}

			Our observable quantity for the galaxy-galaxy lensing measurement is $\Delta \Sigma$,
           which is estimated in bins of $r_p$ as
    		\begin{equation}
            	\widehat{\Delta \Sigma}_{gR}(r_p)=\frac{\sum_{ls}w_{ls}\gamma_t^{(ls)}\Sigma_c^{(ls)}}{\sum_{Rs}w_{Rs}}-
				\frac{\sum_{Rs}w_{Rs}\gamma_t^{(Rs)}\Sigma_c^{(Rs)}}{\sum_{Rs}w_{Rs}}.
        	    \label{eqn:sigma_estimator}
            \end{equation}

            The summation is over all lens-sources (``ls'') pairs. $\gamma_t$ is the tangential shear measured in the
            lens-source frame. $\Sigma_c$ is the geometric factor defined in Eq.~\eqref{eqn:sigma_crit}, and
            the optimal weight $w_{ls}$ for each lens-source pair 
            \referee{($w_{Rs}$ is defined analogously for random-source pairs)} is defined as
            \citep[see][]{Mandelbaum2005}
            \begin{equation}
               w_{ls}=\frac{\Sigma_c^{-2}}{\sigma_\gamma^2+\sigma_{SN}^2}.
            \end{equation}
            The $\Sigma_c^{-2}$ enters the inverse variance weight because we defined the $\Delta\Sigma$ in
            Eq.~\eqref{eqn:sigma_estimator} as the maximum likelihood estimator \citep{Sheldon2004}.
            Note that the denominator in Eq.~\eqref{eqn:sigma_estimator} has a sum over weights $w_{Rs}$, measured by 
            using
            random lenses rather than lens galaxies. Division by $\sum_{Rs} w_{Rs}$ rather than
            $\sum_{ls} w_{ls}$ corrects for the dilution of the shear signal by
            source galaxies that are physically associated with the lens but appear to be 
            behind the lens due to photo-$z$ scatter. These galaxies do not contribute any shear but are counted in
            the total weights (sum over $w_{ls}$). The correction factor for this effect $\sum_{ls}
            w_{ls}/\sum_{Rs} w_{Rs}$ \referee{(properly normalized to account for different number
              of random and real lenses)} is 
            usually called the boost
            factor \citep{Sheldon2004,Mandelbaum2005} and is $\sim1$ for the scales we use in this work 
            $r_p\gtrsim1\mpch$. 
            Finally, we subtract the shear signal measured around the
            random points to remove any systematics that may contribute a spurious shear signal at large
            scales, and to construct a more optimal estimator. % in the spirit of \cite{Landy1993}. 
            Throughout this paper, the subscript `gR' is used to indicate that the measurement
            around random points is 
            subtracted from the measurement around the lenses:
            \begin{equation}
            	\widehat{\Delta \Sigma}_{gR}(r_p)=\widehat{\Delta \Sigma}_{g}(r_p)-\widehat{\Delta \Sigma}_{R}(r_p)
            \end{equation}
            
            One of the main 
            goals of this paper is to test how the subtraction of the signal measured around random
            points impacts the covariance matrix of the
            final measurement. Hence, we will study the signals measured around galaxies and randoms separately as well. 
            We will refer to the signal measured around galaxies by $\widehat{\DS}_g$ and around random
            points by $\widehat{\DS}_R$. The ratio of the 
            number of random points used to the number of lens galaxies is $N_R$:
            \begin{equation}
				N_R=\frac{\text{Number of random lenses}}{\text{Number of lens galaxies}}
			\end{equation}
			In case of $N_R=0$, $\widehat{\DS}_{gR}\equiv\widehat{\DS}_g$. 
    
            To estimate jackknife errors, we use
			100 approximately-equal area ($\sim$10 degrees on a side) jackknife regions to obtain the jackknife mean and
			errors for each $r_p$ bin.        
         
	\subsection{Covariance: theoretical expectations}
		As is derived in appendix~\ref{appendix:covariance}, the covariance for $\DS_g$ is given by
		%\begin{widetext}
			\begin{align}
				&\text{Cov}(\DS_g)(|\vec r_{p,i}|,|\vec r_{p,j}|)=\nonumber\\
				&\left[\frac{\mathcal A_W(\vec r_{p,i}-\vec r_{p,j})}{\mathcal A_W(\vec r_{p,i})
				\mathcal A_W(\vec r_{p,j})}\frac{1}{L_W}
				\int \frac{\mathrm{d} k\,k}{2\pi} J_2(kr_{p,i})J_2(kr_{p,j})\right. \nonumber\\&\left.
				\left(\Sigma_c^2\left(P_{gg}(k)+{N_g}\right)
				\left(P_{\kappa\kappa}(k)+N_\gamma\right)+\Delta\Pi_2\overline\rho P_{g\delta}^2
				+T_{g\gamma g\gamma}\right)
				\right]\nonumber\\&
				+\left\{\frac{1}{\mathcal A_W(\vec r_{p,i})\mathcal A_W(\vec r_{p,j})L_W}
				\int \frac{\mathrm{d} k\,k}{2\pi}J_2(kr_{p,i})J_2(kr_{p,j})\right.\nonumber\\&\left.
				\tilde W(k)^2\Sigma_c^2
				\left(P_{\kappa\kappa}(k)+N_{\gamma}\right)
				\right\}\label{eq:theory_GG_cov}.
			\end{align}
		%\end{widetext}
		
		Here	the lens 	
		galaxy power spectrum can be written as $P_{gg}=b_g^2P_{\delta\delta}(k)$ in the linear bias
        regime, the lens galaxy shot noise 
		power spectrum is $N_g=\frac{1}
		{\overline{n}_g}$, the 
		shape noise term is $N_{\gamma}=\frac{\sigma_{\gamma}^2}{\overline{n}_s}$, the galaxy-shear 
		cross-power spectrum is $P_{g\gamma}=\overline\rho b_gr_{cc} P_{\delta\delta}(k)$, and 
		the convergence power spectrum $P_{\kappa\kappa}$ is given in Eq.~\eqref{eq:convergence}.
		We compute the $\mean{\Sigma_c}$  
		\referee{when performing the measurements,} and use $\mean{\Sigma_c}\sim4.7\times10^3\frac{h\Msun}
		{\text{pc}^2}$, $n_s\sim8h^{2}\text{\mpc}^{-2}$ (after accounting for weights) 
		in theoretical covariance calculations.  
		 $J_2$ is the second order spherical Bessel function, $\sigma_\gamma\sim0.36/2\mathcal R$ 
		is the shape noise,  
        $\Delta\Pi_2\approx700\mpch$ is the  
        line-of-sight integration length using the lensing window function,  
		$W(k)$ is the projected lens window function in 
		Fourier space (see appendix~\ref{appendix:covariance} for the expressions for the window function) and 
		$L_W$ is the line-of-sight length of the lens window function.
		$T_{g\gamma g\gamma}$ is the connected part of the covariance, which we will ignore in numerical calculations.
		$\mathcal A_W$ (defined in Eq.~\ref{eq:AW_numerical}) 
		is the window function-dependent effective area covered by each bin, 
		and accounts for the edge effects due to the survey window. For 
		scales much smaller than the survey window, $\mathcal A_W\approx A_W$, where $A_W$ is the survey area. 
		In the $\DS$ measurements in this work, 
		we only divide the lens sample into jackknife regions, but the source sample stays the same and 
		hence the edge effects are small. When calculating numerical predictions for the jackknife
        errors, we set the window function 
		$\mathcal A_W\approx A_W$ (ideally 
		we should set $\mathcal A_W$ for jackknife to be same as that of full sample, but in the
        case of an idealized LOWZ-sized 
		window, $\mathcal A_W\approx A_W$ for the scales of interest). 
		In appendix~\ref{appendix:clustering}, we show the effects of $\mathcal A_W$ on the jackknife covariance 
		in the case of clustering measurements.
		
		The covariance for $\DS_{gR}$ is similar to what is shown in Eq.~\eqref{eq:theory_GG_cov},
        except that it does not contain the last 
		term in curly brackets, $\{\}$. This term arises because of the non-zero mean value of 
		the lens density (here the lens sample is assumed to be normalized and hence its mean is 1). 
		This term is independent of the lens over-density
		and only depends on the window function of the lens sample. Hence it get removed when the measurement around 
		random points is subtracted from the measurement around galaxies.
         
	\subsection{Covariance matrix estimation methods}\label{subsec:covest}

         To estimate the covariance matrix, we use two different methods. The first is the jackknife method, in which we divide
         the whole survey into $N_\text{Jk}=100$ approximately equal-area regions \referee{($\sim90$
           degrees$^2$ $\equiv 76^2,125^2, 163^2$ [Mpc/h]$^2$ at $z=$ 0.16, 0.27, and 0.36
           respectively)}.  We then make $N_\text{Jk}$ measurements by dropping 
         one region at a time, so that each measurement contains $N_\text{Jk}-1$ regions. The jackknife
         variance estimate (diagonals of the covariance matrix) is then
         \begin{equation}
         	\text{Var}_\text{Jk}(\widehat\DS)=\frac{N_\text{Jk}-1}{N_\text{Jk}}\sum_{i=1}^{N_\text{Jk}}(\DS_i-\overline
			\DS)^2
         \end{equation}

         Our second method is to measure \DS\ using $N_M$ mock realizations of the lens sample and then 
         compute the standard deviation (``Std'') of the measurement across all realizations:
         \begin{equation}
         	\text{Var}_\text{Std}(\widehat\DS)=\frac{1}{N_M-1}\sum_{i=1}^{N_M}(\DS_i-\overline\DS)^2
         \end{equation}
		
		Finally, for comparison, we also show error estimates using subsamples of the survey. We use the same
		subsampling as in the jackknife method, but in this case we perform the measurements using one
        subsample at a time. The variance in this case (error on
        the mean) is 
         \begin{equation}
         	\text{Var}_\text{subsample}(\widehat\DS)=\frac{1}{N_{\text{Jk}}(N_\text{Jk}-1)}\sum_{i=1}^{N_\text{Jk}}
			(\DS_i-\overline\DS)^2
         \end{equation}

\section{Data}\label{sec:data}
	\subsection{SDSS}
		 The SDSS \citep{2000AJ....120.1579Y} imaged roughly $\pi$ steradians
		of the sky, and the SDSS-I and II surveys followed up approximately one million of the detected
		objects spectroscopically \citep{2001AJ....122.2267E,
		  2002AJ....123.2945R,2002AJ....124.1810S}. The imaging was carried
		out by drift-scanning the sky in photometric conditions
		\citep{2001AJ....122.2129H, 2004AN....325..583I}, in five bands
		($ugriz$) \citep{1996AJ....111.1748F, 2002AJ....123.2121S} using a
		specially-designed wide-field camera
		\citep{1998AJ....116.3040G} on the SDSS Telescope \citep{Gunn2006}. These imaging
		data were used to create
		the  catalogues of shear estimates that we use in this paper.  All of
		the data were processed by completely automated pipelines that detect
		and measure photometric properties of objects, and astrometrically
		calibrate the data \citep{Lupton2001,
		  2003AJ....125.1559P,2006AN....327..821T}. The SDSS-I/II imaging
		surveys were completed with a seventh data release
		\citep{2009ApJS..182..543A}, though this work will rely as well on an
		improved data reduction pipeline that was part of the eighth data
		release, from SDSS-III \citep{2011ApJS..193...29A}; and an improved
		photometric calibration \citep[`ubercalibration',][]{2008ApJ...674.1217P}.

	\subsection{SDSS-III BOSS}
			Based on the SDSS photometric catalog,  galaxies were selected for spectroscopic
			observation
			\citep{Dawson:2013}, and the BOSS spectroscopic survey was performed
			\citep{Ahn:2012} using the BOSS spectrographs \citep{Smee:2013}. Targets
			were assigned to tiles of diameter $3^\circ$ using an adaptive tiling
			algorithm \citep{Blanton:2003}, and the data were processed by an
			automated spectral classification, redshift determination, and parameter
			measurement pipeline \citep{Bolton:2012}.

			We use SDSS-III BOSS data release 12 \citep[DR12;][]{Alam2015,Reid2016}
			LOWZ galaxies in the redshift range $0.16<z<0.36$. 
			%\citep[see ][ for details of sample selection]{Singh2016}.
			The LOWZ sample consists of Luminous Red Galaxies (LRGs) at $z<0.4$, selected
			from the SDSS DR8 imaging data and observed
			spectroscopically in the BOSS survey. The sample is approximately volume-limited in the
			redshift range $0.16<z<0.36$, with a number
			density of $\bar{n}\sim 3\times10^{-4}~h^3\text{Mpc}^{-3}$ \citep{Manera2015,Reid2016}. 
			We use the same sample as used by \cite{Singh2016}, who mask out certain regions on the sky which have
			higher galactic extinction or poor imaging quality \citep{Reyes2012}, which leaves 225,181 galaxies 
			in the sample.

	\subsection{Re-Gaussianization Shapes and Photometric redshifts}
		The shape measurements for the source sample used in this work
		are described in more detail in \cite{Reyes2012}. Briefly, these shapes are measured
		using the re-Gaussianization technique developed by \cite{Hirata2003}. The
		algorithm is a modified version of ones that use ``adaptive moments'' (equivalent to fitting
        the light intensity profile to an elliptical Gaussian), determining shapes of the
        PSF-convolved galaxy image based on adaptive moments and then correcting the resulting
        shapes based on adaptive moments of the PSF.   The re-Gaussianization method involves
        additional steps to correct for  non-Gaussianity of both the PSF and the galaxy surface
        brightness profiles \citep{Hirata2003}. The components of the distortion are defined as
		\begin{equation}\label{eqn:distortion}
			(e_+,e_\times)=\frac{1-(b/a)^2}{1+(b/a)^2}(\cos 2\phi,\sin 2\phi),
		\end{equation}
		where $b/a$ is the minor-to-major axis ratio and $\phi$ is the position angle of the major
		axis on the sky with respect to the RA-Dec coordinate system. The ensemble average of the
        distortion is related to the shear as
		\begin{align}
			\gamma_+,\gamma_\times&=\frac{\langle e_+,e_\times\rangle}{2\mathcal
			R}\label{eqn:regauss_shear}\\
			\mathcal R&=1-\frac{1}{2}\langle e_{+,i}^2+e_{\times,i}^2-2\sigma_i^2\rangle\label{eq:R}
		\end{align}
		where $\sigma_i$ is the per-component measurement uncertainty of the galaxy distortion, and
		${\mathcal R\approx0.87}$ is the shear responsivity representing the response of an ensemble of
        galaxies with some intrinsic distribution of distortion values
		to a small shear \citep{Kaiser1995,Bernstein2002}. \referee{A discussion of corrections for shear-related
          systematic biases and the residual systematic uncertainties can be found in
          \cite{Mandelbaum2013}.  These estimates are based on a combination of null tests using the
        shear catalog and external image simulations.}
		
		The photometric redshifts for the catalog were estimated using the template-fitting code ZEBRA \citep{Feldman2006}.
		Using photometric redshifts for the source sample introduces a bias in galaxy-galaxy lensing through 
		misestimation of the $\Sigma_c$ factor (with the most severe misestimation arising due to the
        inclusion of some lens-foreground ``source'' pairs due to scatter in 
		photometric redshifts). 
		\cite{Nakajima2012} showed that this bias can be large, but can be estimated to within $2\%$ 
		using a representative calibration sample with spectroscopic redshifts. We compute these correction factors 
		using the method of \cite{Nakajima2012} with the LOWZ lens redshift distributions to be $\sim10\%$ and 
		then multiply our measurements with  a calibration factor of 1.1.

	\subsection{Mock source catalog}

		We generate 100 mock catalogs of the shape sample by randomly rotating the shapes of galaxies in the real 
		source sample, while keeping their positions (RA, Dec, $z$) fixed. Random rotations remove any coherent shear 
		(cosmological or due to systematics) in the sample while maintaining the shape noise and measurement noise 
		in each realization. As a result, $\DS$ measurements using rotated (mock) sources will not have any coherent 
		signal and their 
		covariance matrix will only have contributions from shape noise and measurement noise. The comparison of the 
		covariance matrix of mocks with the covariance from real sources will allow us to study the contribution of 
		shape noise and measurement noise to the covariance in the real data.
		
	\subsection{QPM mocks}
		To estimate the galaxy-galaxy lensing covariance matrix using a mock lens sample, we use the QPM mocks \citep{White2014} which have been
		used in several BOSS analyses \citep[e.g.,][]{Cuesta2016,Grieb2016,Marin2016}. QPM mocks are constructed using 
		the quick particle mesh method \citep{White2014} to mimic the large-scale clustering properties of BOSS 
		galaxies. In this work, we use 100 QPM mocks with the same sky coverage, mask and jackknife splitting as in
		the LOWZ sample.

\section{Results}\label{sec:results}
	In this section we present our results from measuring galaxy-galaxy lensing using different estimators and different 
	combinations of lens and source galaxies. We perform several tests to study the effects on the estimated covariance 
	by using different covariance estimation methods, 
	varying $N_R$, varying the clustering properties of lens sample, and varying the source sample
    (without and with systematics). A summary of the various terms in the covariance that contribute for different 
    combinations of lens and source samples is presented in 
    Table~\ref{tab:summary}, and a summary of the results is in Figure~\ref{fig:summary}.

   \subsection{LOWZ lensing results}\label{ssec:results_lowz}
\begin{figure}
            \centering
            \includegraphics[width=\columnwidth]{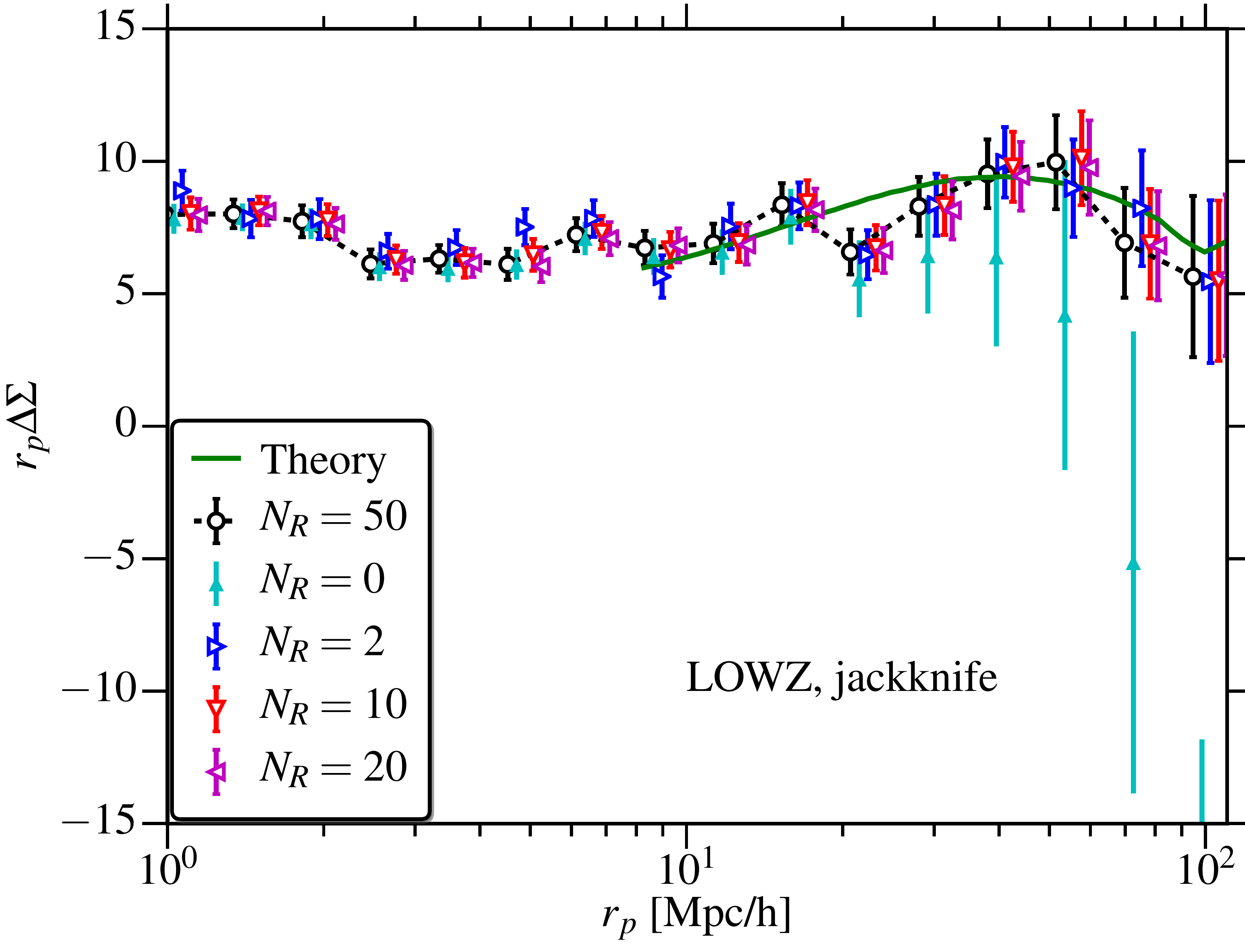}
            \caption{\DS\ measured for the LOWZ sample, with different numbers of random catalogs
              used. The errors shown are from the jackknife method. The signal without subtraction of the lensing signal around random 
            	points ($N_R=0$, cyan points) shows the presence of additive systematics in the SDSS source 
				sample. These systematics are removed with the subtraction of the signal measured
                around random points. Theory 
				predictions use the linear theory$+$halofit power spectrum with fixed cosmology along with the
                best-fitting linear bias and $r_\text{cc}=1$, and fitting
				was done for $10\mpch<r_p<65\mpch$. 
				}
         \label{fig:lowz_DS}
      \end{figure}

		We begin by showing the galaxy-galaxy lensing measurements using the LOWZ lens sample.
		Fig.~\ref{fig:lowz_DS} shows $\DS$ measured using LOWZ lens galaxies and different numbers
        of random points. 
		When using
		no randoms (the $N_R=0$ case), there is evidence for a spurious systematic signal at large scales.
		\referee{This spurious signal arises because the PSF correction method used to measure the
          galaxy shapes is unable to fully
          remove all of the PSF anisotropy.  The SDSS survey strategy results in large-scale
          coherent PSF anisotropy which, when improperly removed, causes a large-scale coherent
          galaxy shape alignment \citep[see][for a 
		detailed discussion]{Mandelbaum2005,Mandelbaum2013}}. 
		The fact that this spurious signal gets removed when the measurement around randoms is subtracted has been the 
		primary motivation for the subtraction of the signal around random points in SDSS
        galaxy-galaxy lensing measurements.
      
     \begin{figure}
            \centering
            \includegraphics[width=\columnwidth]{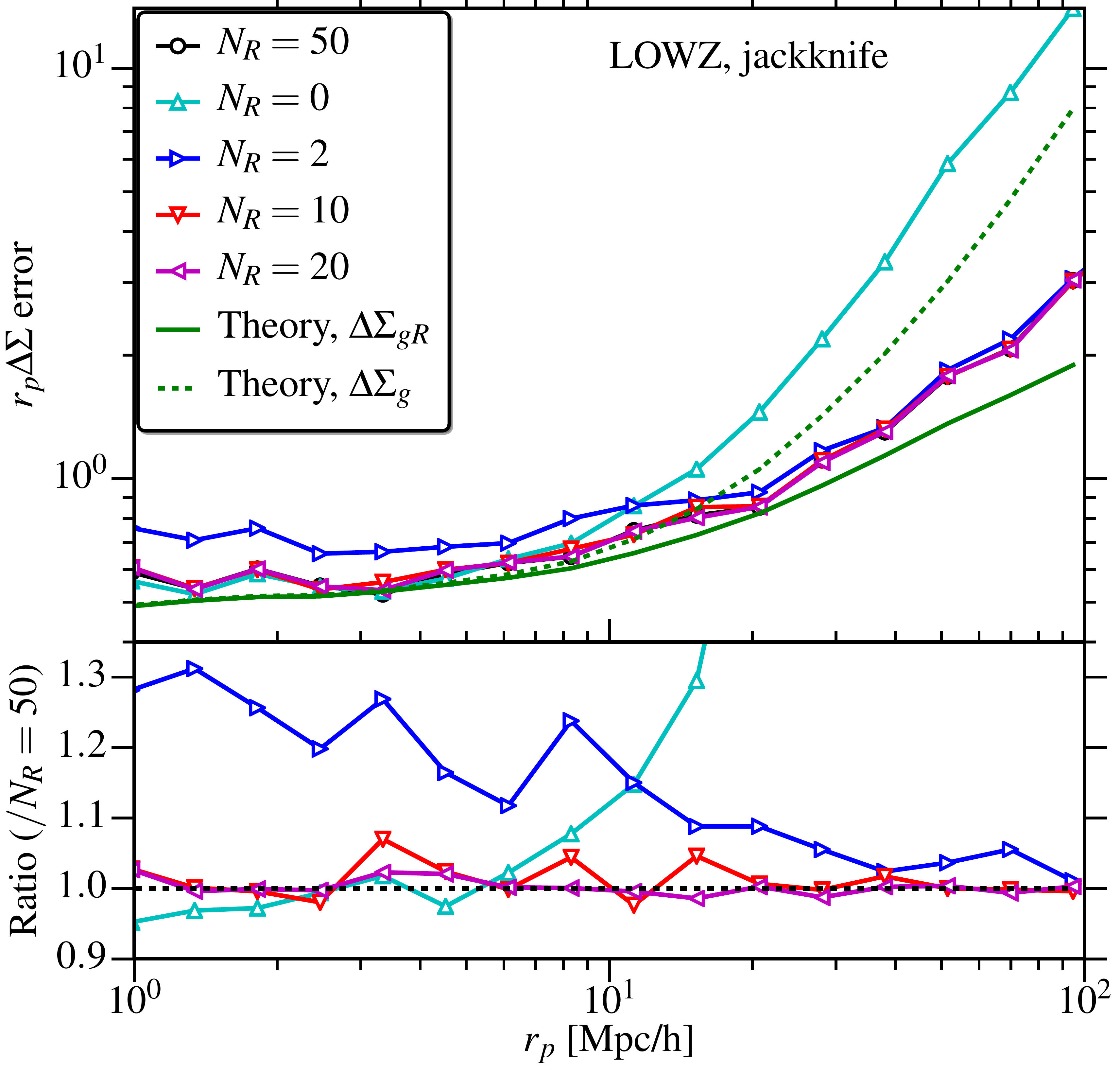}
            \caption{Jackknife errors in the LOWZ \DS\ measurement (square root of diagonal elements of the covariance 
            	matrix), 
            	for different numbers of 
	            random catalogs. The errors generally follow the $\propto 1/r_p$ scaling expected
                from shape noise and logarithmic binning in $r_p$,
	            though there is some saturation at large scales due to the correlated shape noise and systematics. 
	            Also shown are the theory predictions, which are consistent with data at small scales, though there 
	            are differences at large scales due to systematics that are not included in the
                theory.
                }
            \label{fig:lowz_error}
      \end{figure}

       However, the subtraction of the signal around random points also reduces the 
		errors in the measurements, especially at large scales (the noise in the $\DS_g$ term has contributions from 
		systematics as well as shape noise, as we will show in later sections.). 
          In Fig.~\ref{fig:lowz_error}, we show the variations in 
		the error estimates (square root of the diagonal covariance matrix elements) with different numbers of random 
		catalogs $N_R$. 
		At small scales, where the errors follow the expected scaling for shape noise 
		($\propto 1/r_p$ in logarithmic $r_p$ bins), subtracting the signal around random points increases the error 
		estimates, though with $N_R\gtrsim10$, the errors converge to $N_R=0$ case. The errors in this regime
        should scale with $N_R$ as
		\begin{equation}
			\left(\frac{\delta\DS(N_R)}{\delta\DS(N_R=0)}\right)^2=1+\frac{1}{N_R}.
		\end{equation}
		Given that the jackknife error estimates using 100 regions have uncertainty of order 
		$\sim15\%$ \citep[$\sqrt{2/99}$;][]{Taylor2013}, using $N_R=10$ is sufficient and henceforth our results will 
		use $N_R=10$ unless a different value is explicitly given. However, note that when using large numbers of mocks for error 
		estimates, more randoms might be required. 
 
 		At large scales, contributions to the noise from systematics and the correlated shape noise 
		($P_{gg}(P_{\gamma\gamma}+N_{\gamma\gamma})$ term) start dominating and hence the error estimates diverge from 
		the $1/r_p$ scaling.
		The errors are mostly consistent with the theoretical predictions calculated using Eq.~\eqref{eq:theory_GG_cov}. 
		At large scales there is a contribution from the systematics that is not included in the theory 
		predictions, hence the errors diverge from those predictions especially for the $N_R=0$ case where
        systematics are most important. 
		We distinguish between the different terms in the variance in the following sections.
    
\begin{figure*}
      \centering
      \begin{subfigure}{\columnwidth}
	      \includegraphics[width=\columnwidth]{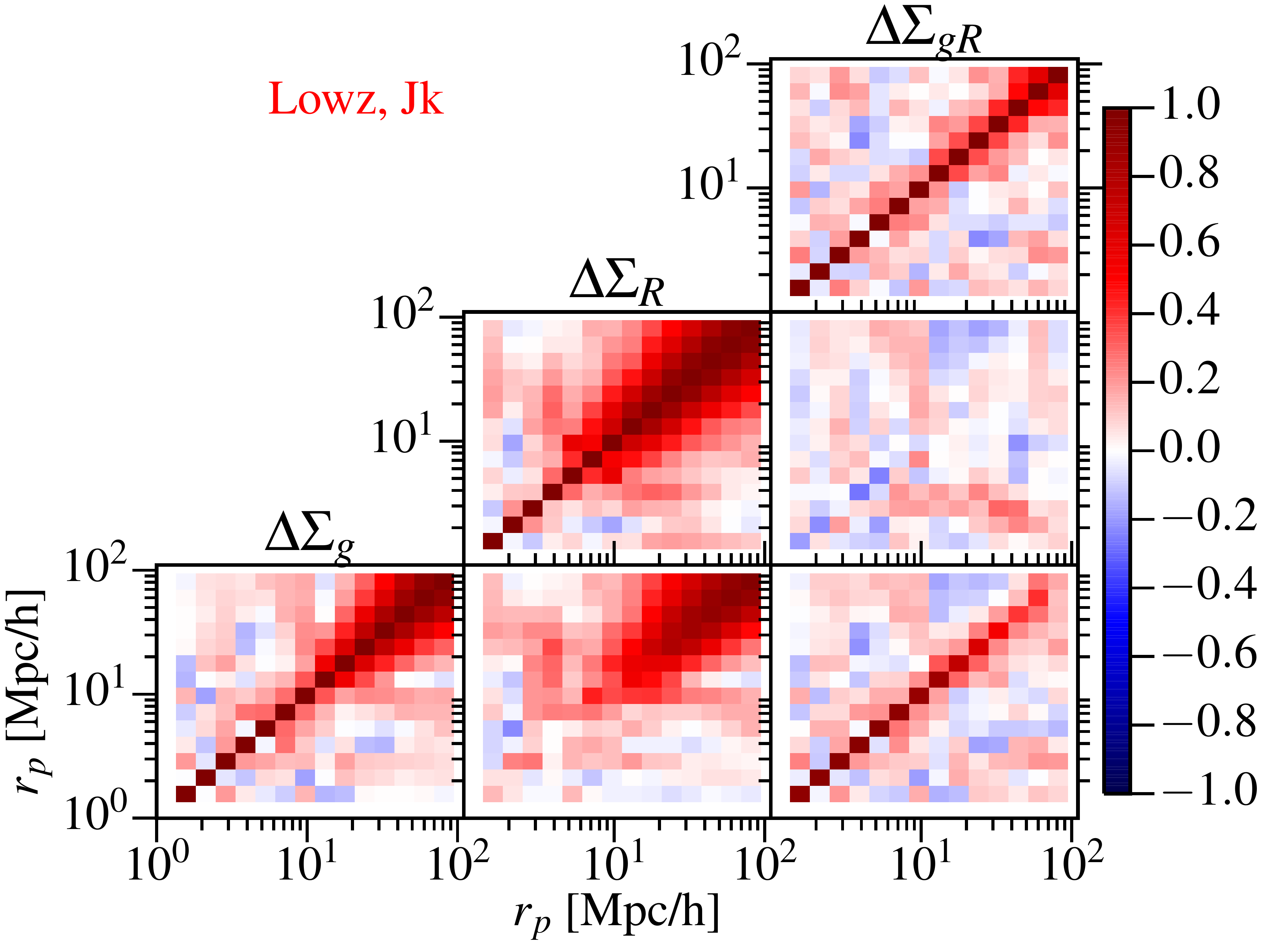}
	   \end{subfigure}
	   \begin{subfigure}{\columnwidth}
	      \includegraphics[width=\columnwidth]{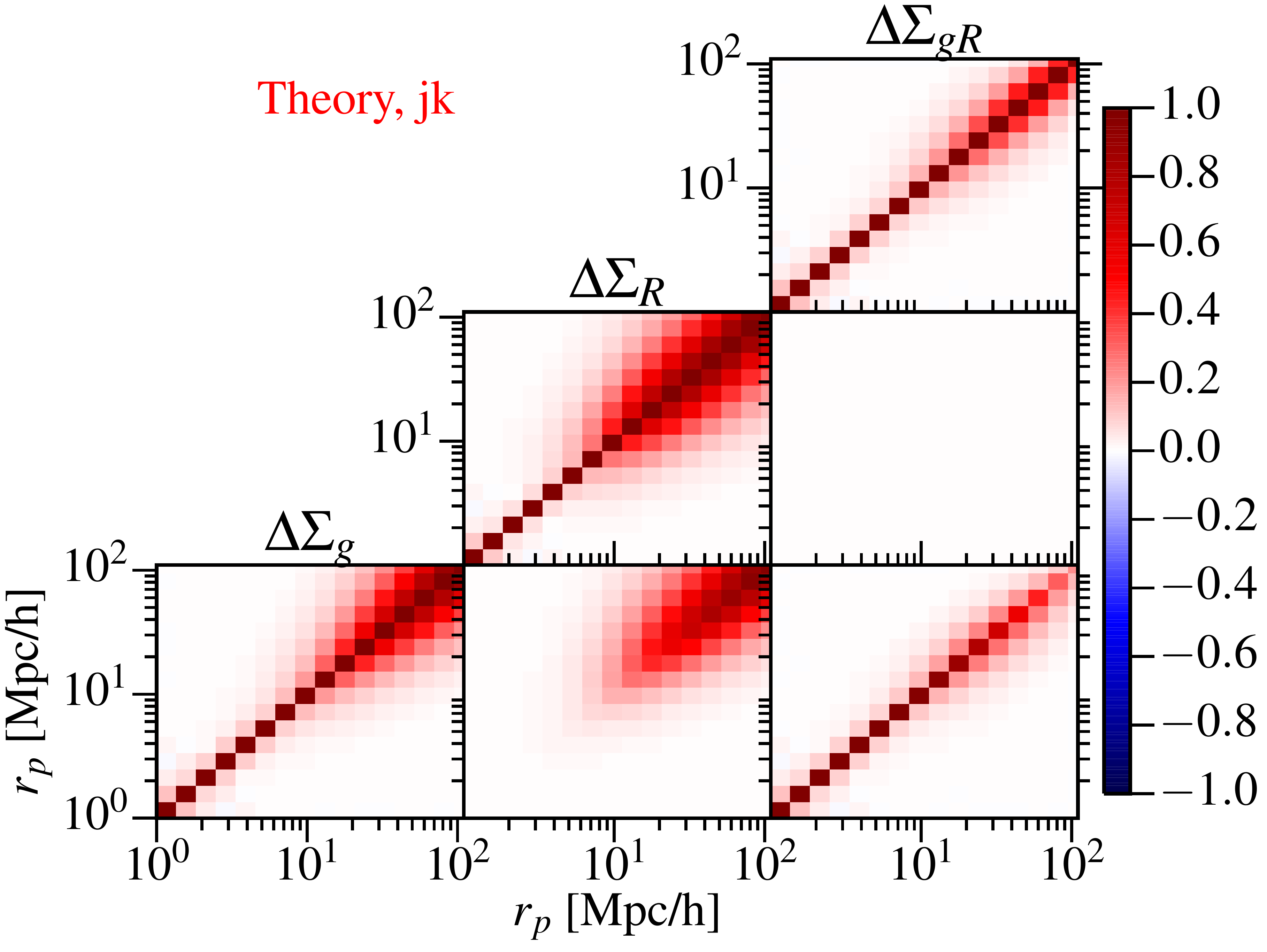}
	   \end{subfigure}
	   \begin{subfigure}{\columnwidth}
	      \includegraphics[width=\columnwidth]{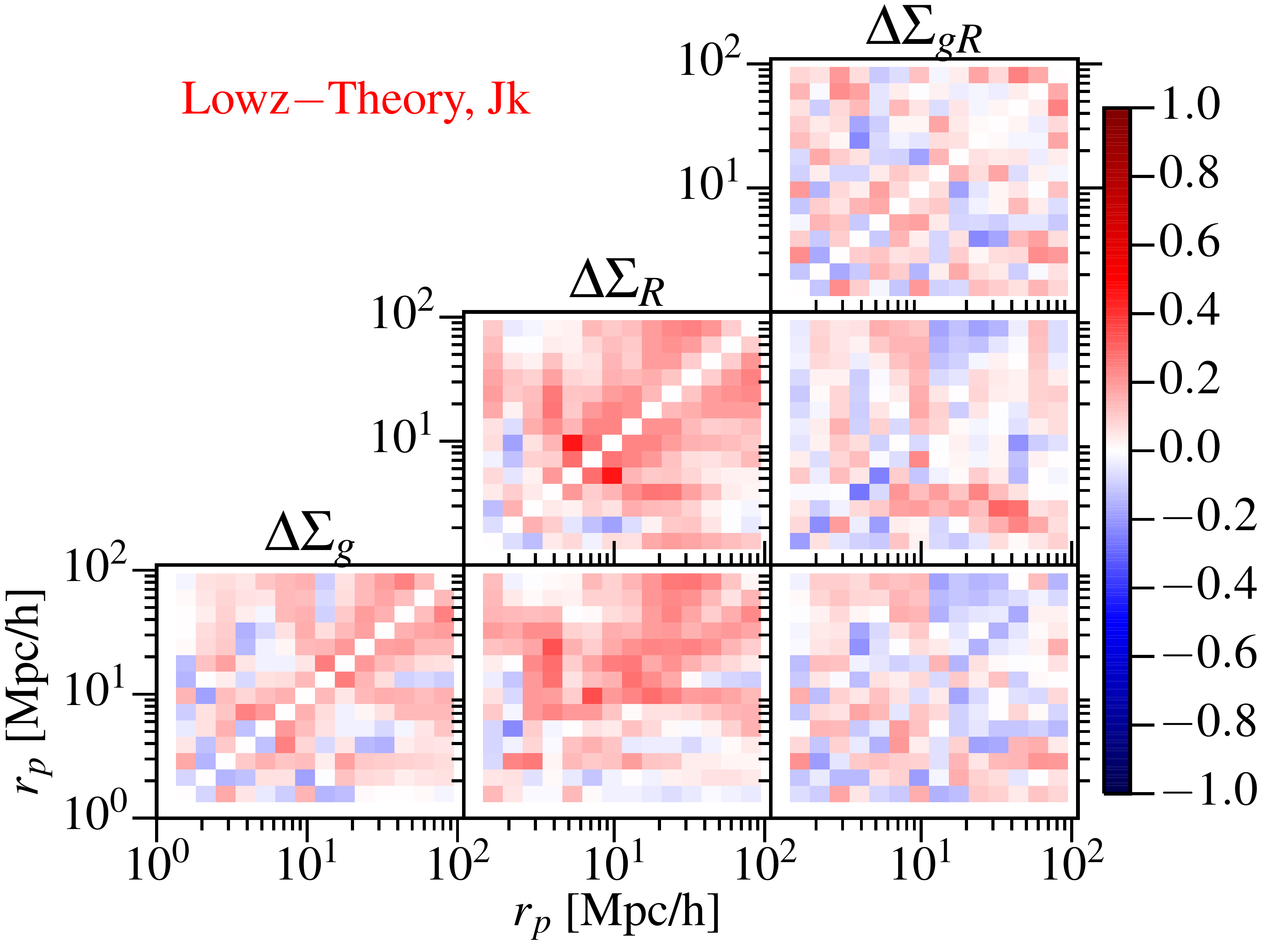}
	   \end{subfigure}
	   	   \begin{subfigure}{\columnwidth}
	      \includegraphics[width=\columnwidth]{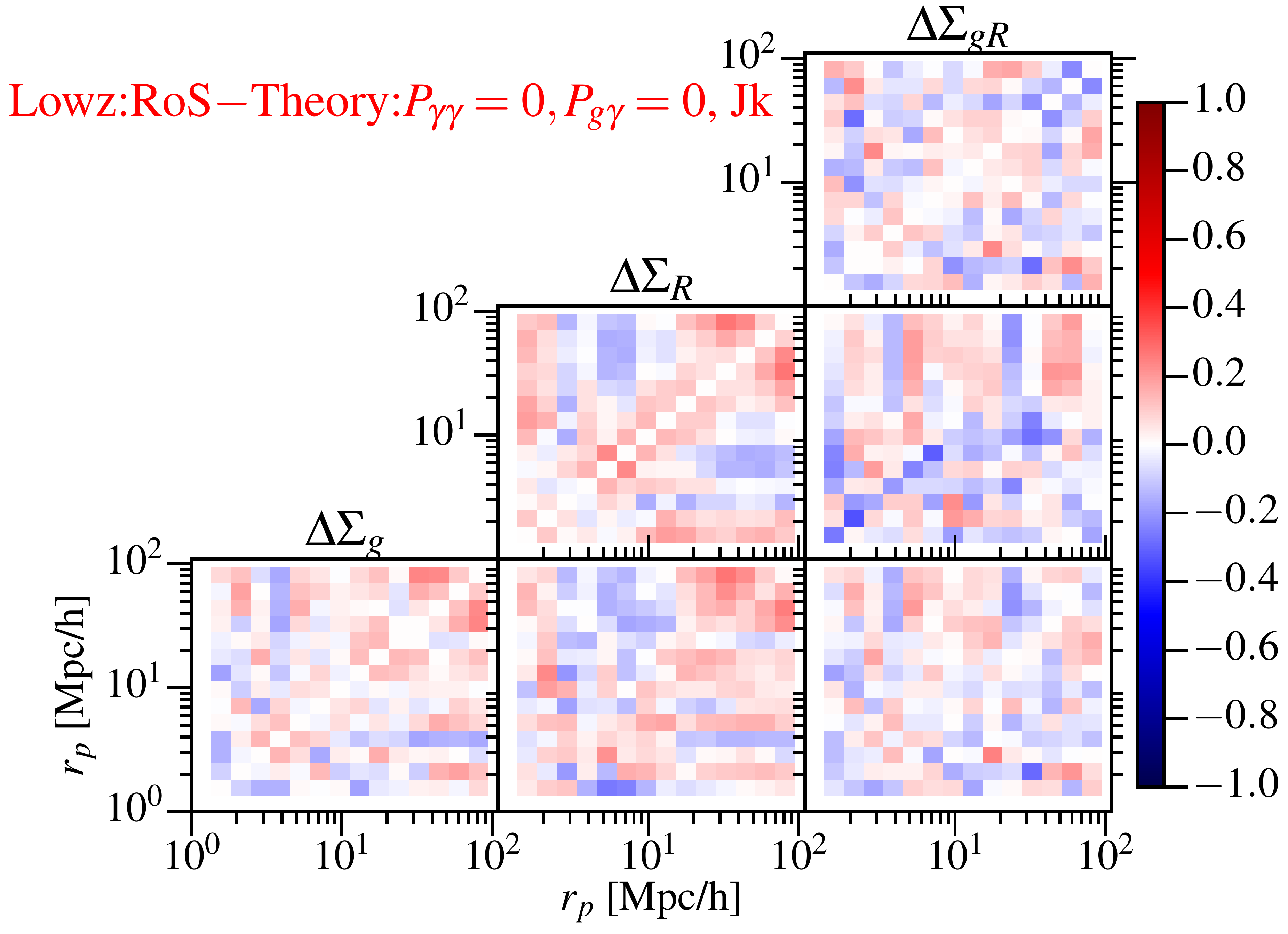}
	   \end{subfigure}
      \caption{\emph{Upper row}: 
      	Correlation and cross-correlation matrices for \DS\ measured around galaxies ($\DS_g$), randoms ($\DS_R$) 
	      and the difference of the two ($\DS_{gR}$), both from theory and data (jackknife).
	      Due to the shape noise, there are strong auto- and cross-correlations 
	      at large scales in both $\DS_g$ and $\DS_R$. Subtracting the measurement around the random points removes most of the 
	      correlated noise (from systematics and correlated shape noise), 
	      though there are still some residual bin-to-bin correlations in $\DS_{gR}$, primarily 
	      due to the clustering of lens galaxies.
	      \emph{Bottom row}: 
	      The difference between the correlation matrices (note we do not take the difference of the covariances here)
	      from the jackknife and the theoretical predictions (\emph{left}) and from the mocks and
          the theoretical predictions 
	      (\emph{right}). In the case of the jackknife matrices, there are systematics that are not
          included in the theory predictions, so the latter are under-predicated. In the case of the
          mock sources \referee{(or rotated sources, RoS)}, shear systematics are removed and the 
	      theory predictions are consistent with the data within the noise in the jackknife covariances.
    }
      \label{fig:lowz_corr}
    \end{figure*}

%		In case of our measurement, the additive systematics also further increase the large scale noise. Fortunately, 
%		the correlated shape noise also similarly affects the measurement around randoms and as shown in 
		Fig.~\ref{fig:lowz_corr} shows the correlation and cross-correlation matrices for $\DS_g$, $\DS_R$ and $\DS_{gR}
		$, both from theory and data and their difference. 
		The measurements of $\DS$ around galaxies ($\DS_g$) and randoms ($\DS_R$) are highly correlated for $r_p\gtrsim10\mpch$. 
		When we subtract the measurement around randoms, this 
		correlated noise gets removed and hence the noise in $\DS_{gR}$ decreases compared to that
        in $\DS_g$  at large scales. The bin-to-bin correlations 
		also decrease, though there are still some residual correlations due to the clustering 
		of the lens sample and the effects of systematics. Since the theory prediction does not include systematics, the residuals after 
		subtracting the theory correlation matrix from the jackknife are not consistent with zero.
		A cleaner test of the theoretical expressions will use randomly rotated sources, which do not have any
        systematic shear correlations, in the next subsection.
		
	\subsection{Mock Sources}\label{ssec:results_source_rotation}
   		In this section we quantify the effects of additive shear systematics on the covariance 
		estimation, especially on the differences in the errors with and without $\DS_R$ subtracted. 
		We create 100 mock realizations of the source sample by randomly rotating the source
        galaxies. The resulting source catalogs should exhibit no coherent signals of cosmological
        origin or due to systematics.  When measuring $\Delta\Sigma$ around the LOWZ galaxies with these
        randomly rotated source catalogs, we should observe realistic levels of correlated shape
        noise, but no systematics, cosmic variance, or super-sample covariance.  All terms involving shear 
        correlations -- $P_{\gamma\gamma}$, $P_{g\gamma}$, and $T_{g\gamma g\gamma}$ -- 
        are zero and hence do not contribute to the covariance.
		
   		\begin{figure}
    	  \centering
	      \includegraphics[width=\columnwidth]{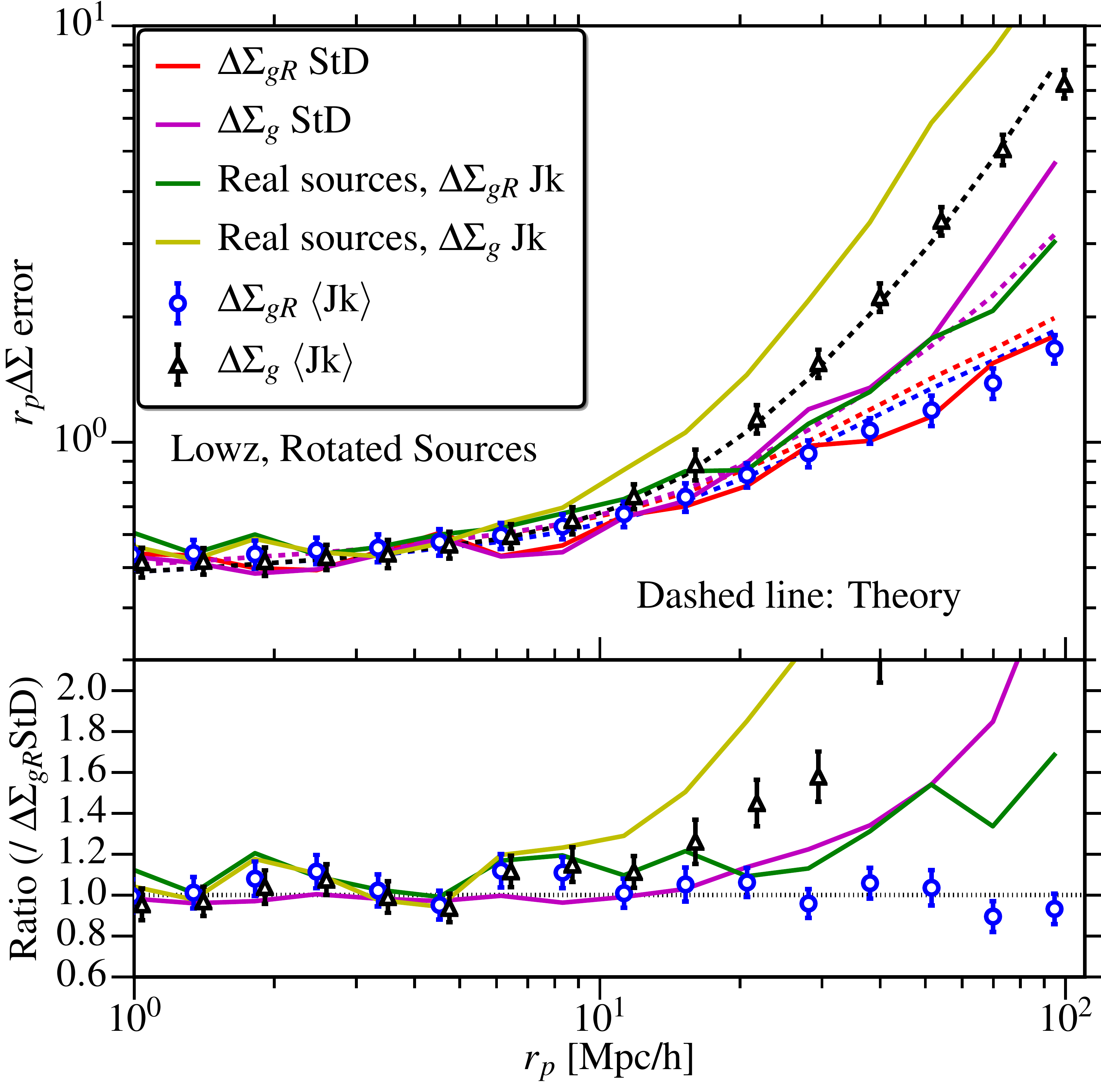}
    	  \caption{$\DS$ errors using different estimators with LOWZ galaxies as lenses and 100 mock realizations 
	  		of the source sample obtained by randomly rotating the SDSS source galaxies. 
			Open black and blue markers show the mean and standard deviation of the jackknife errors. Solid
            red and magenta lines show the errors from the standard deviation (``StD'') 	
            across different realizations of $\DS_g$ and 
			$\DS_{gR}$. Dashed lines are the theory predictions for the curves with corresponding colors.
			In the bottom panel we plot the ratio of the different errors with respect to the StD 
			errors of $\DS_{gR}$. Jackknife errors for the real LOWZ sample are also plotted for comparison ($N_R=10$ 
			for LOWZ). Note that in the $\DS_g$ case, the correlated noise in these mocks is lower than when using real 
			sources due to the removal of the contribution from systematics. 
    		}
	      \label{fig:SR_error}
    	\end{figure}
	
		Fig.~\ref{fig:SR_error} shows the jackknife and standard deviation errors obtained with and without subtracting 
		$\DS_R$. Subtracting $\DS_R$ reduces the errors, and the results are consistent with the theory predictions.
		However, the magnitude of the difference in errorbars for $\DS_{gR}$ vs.\ $\DS_g$  shown
        here  (factor of $\sim2$ at the largest scale) is lower compared to what was seen with real
        SDSS sources 
		(factor of $\sim5$), which suggests that a bit more than half the contribution to the
        errorbars for $\DS_g$ with real sources came from shear correlations, mostly caused by the
        systematics rather than cosmic shear given the low redshift of this sample. 
        In the case of $\DS_{gR}$, the errors computed using the standard deviation (Std) across the realizations 
		are consistent with the jackknife errors with $\DS_R$ subtraction. In the case of $\DS_g$,
        the Std errors are 
		lower than the jackknife errors because of the much larger effective window for Std
        \referee{(full survey window)} compared to the jackknife 
		\referee{(1/$N_{jk}$ of the survey window)}.
	
		As shown in Figs.~\ref{fig:lowz_corr} and~\ref{fig:SR_error}, in the case of the jackknife window, 
        the predictions from theory are consistent with the data at the $\sim10\%$ level for both $\DS_g$ and 
        $\DS_{gR}$,  
        which is within the noise in the jackknife errors. In the case of $\DS_{g}$ with the full
        survey window, the theory 
        predictions for the errors are lower than the jackknife errors. This is likely because when computing the theory predictions, we assume an 
        idealized 
        geometry (see Appendix~\ref{appendix:covariance}), which underestimates the window function
        effects from the  
        realistically complicated window in the data.
        
        Fig.~\ref{fig:SR_error} and the theoretical predictions
        demonstrate that even in a survey with no known additive systematic errors, measurements of
        $\Delta\Sigma$ in the $r_p$ range where correlated shape noise is important will have
        substantially better $S/N$ when using the more optimal $\DS_{gR}$ estimator.
		
%		\begin{figure}
%		      \centering
%		      \includegraphics[width=\columnwidth]{lowz-rand10_SR_corr}
%      		\caption{Similar to Fig.~\ref{fig:lowz_corr}, using $\DS_g$ measured with LOWZ galaxies and real source 
%				galaxies and $\DS_R$ measured using randoms ($N_R=10$) with random sources (randomly rotated) as well.
%				Once the sources are rotated, shape noise is no longer correlated between the $\DS_g$ and $\DS_R$. In 
%				this case, subtracting the $\DS_R$ term does not improve the covariance at the large scales.
%		    }
%      		\label{fig:lowz_SR_corr}
%    	\end{figure}
%		Fig.~\ref{fig:lowz_SR_corr} shows the correlation matrix obtained after $\DS_R$ obtained from random lens 
%		($N_R=10$) using a mock source realization is subtracted from $\DS_g$ measured using LOWZ galaxies with the real 
%		source sample. Since sources are rotated in $\DS_R$, the shape noise is no longer correlated between $\DS_g$ 
%		and $\DS_R$, i.e., $\mean{\DS_{N,g}\DS_{N,R}}=0$ in Eq.~\eqref{eq:DS_error}. In this case subtracting out 
%		$\DS_R$ 
%		does not help and the final errors (shown in fig.~\ref{fig:summary}) are somewhat 
%		larger compared to the no $\DS_R$ subtraction case, 
%		due to the noise contribution from $\DS_R$. 

   \subsection{Lens Mocks}
   	In this section, we vary the lens properties to examine how the covariance depends on the lens
    sample properties.
	
	   \subsubsection{QPM mocks}\label{ssec:results_qpm}
       		In this section we measure $\DS$ around the galaxies in the QPM mocks using the real and mock source
            sample. In both cases shear-galaxy correlations will be absent, i.e., $P_{g\gamma}=T_{g\gamma g\gamma}=0$,
            while the former will include $P_{\gamma\gamma}$ terms
            and the latter will not.
            Even though the QPM mocks have somewhat different clustering at small scales than the real 
            LOWZ sample (see Fig.~\ref{fig:qpm_wgg}), the typical separation between galaxies ($\sim1$--$2$ \mpch) is 
            very similar between the mocks and the LOWZ 
            sample and hence the QPM mocks are adequate to test the effects of lens clustering, $P_{gg}$, 
            on the galaxy-galaxy lensing 
            covariance. The signal around the QPM mocks should include large-scale systematics and realistic levels of
            shape noise.  %There should be no contributions from the connected part of the covariance, 
            %$T_{g\gamma g\gamma}=0$.
    %
    %	   \begin{figure*}
    %	   		\begin{subfigure}{\columnwidth}
    %		    	 \centering
    %		      	\includegraphics[width=\columnwidth]{qpm_DS_error}
    %		    \end{subfigure}
    %	   		\begin{subfigure}{\columnwidth}
    %		    	 \centering
    %		      	\includegraphics[width=\columnwidth]{qpm_SR_DS_error}
    %		    \end{subfigure}
    %    	  \caption{Same as Fig.~\ref{fig:randoms_error}, but now using QPM mocks (200 realizations) as lens galaxies. 
    %    		}
    %	      \label{fig:qpm_error}
    %    	\end{figure*}
    	
\begin{figure*}
	   		\begin{subfigure}{\columnwidth}
	    	  \centering
		      \includegraphics[width=\columnwidth]{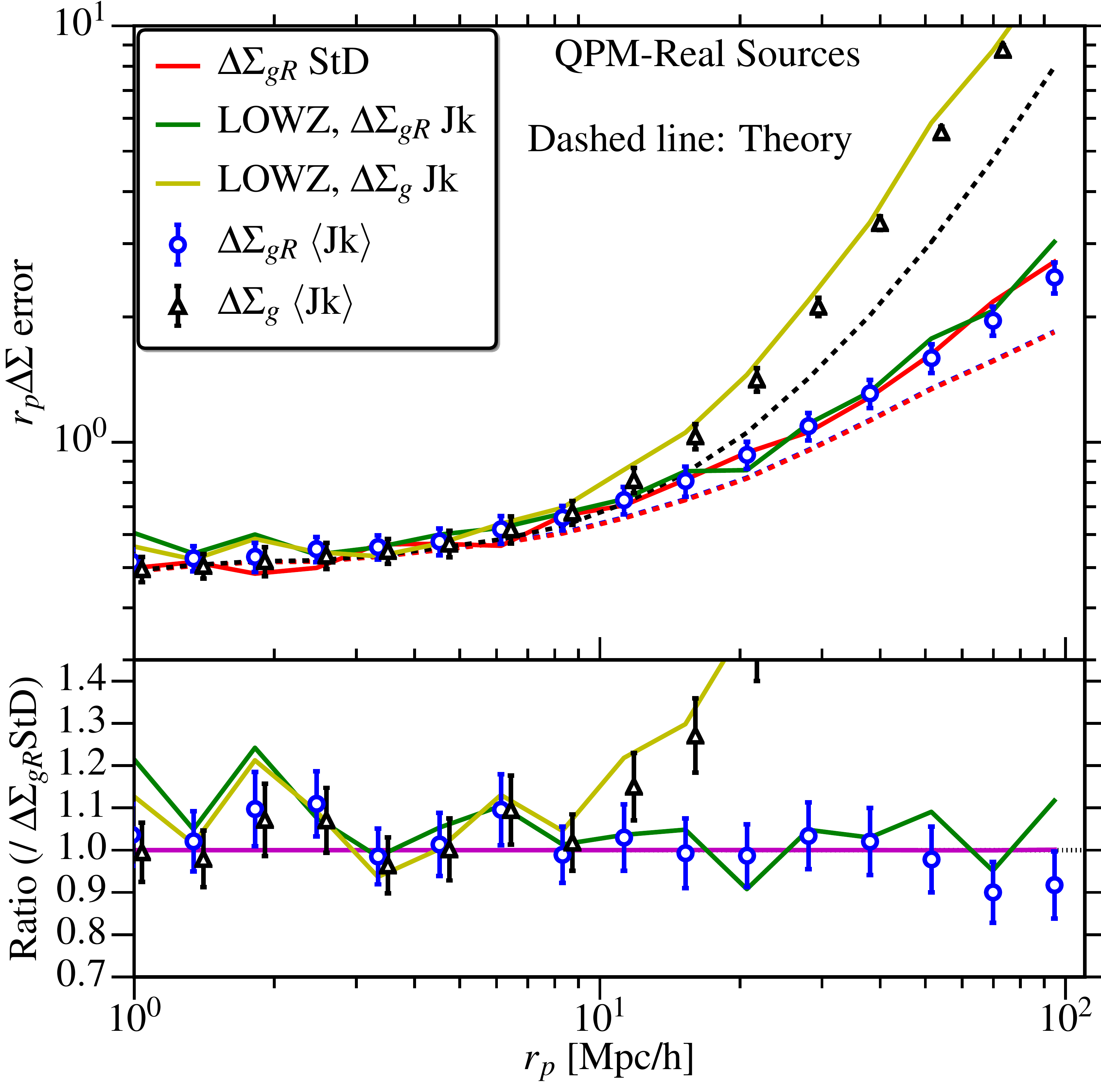}
		    \end{subfigure}
		   	\begin{subfigure}{\columnwidth}
	    	  \centering
		      \includegraphics[width=\columnwidth]{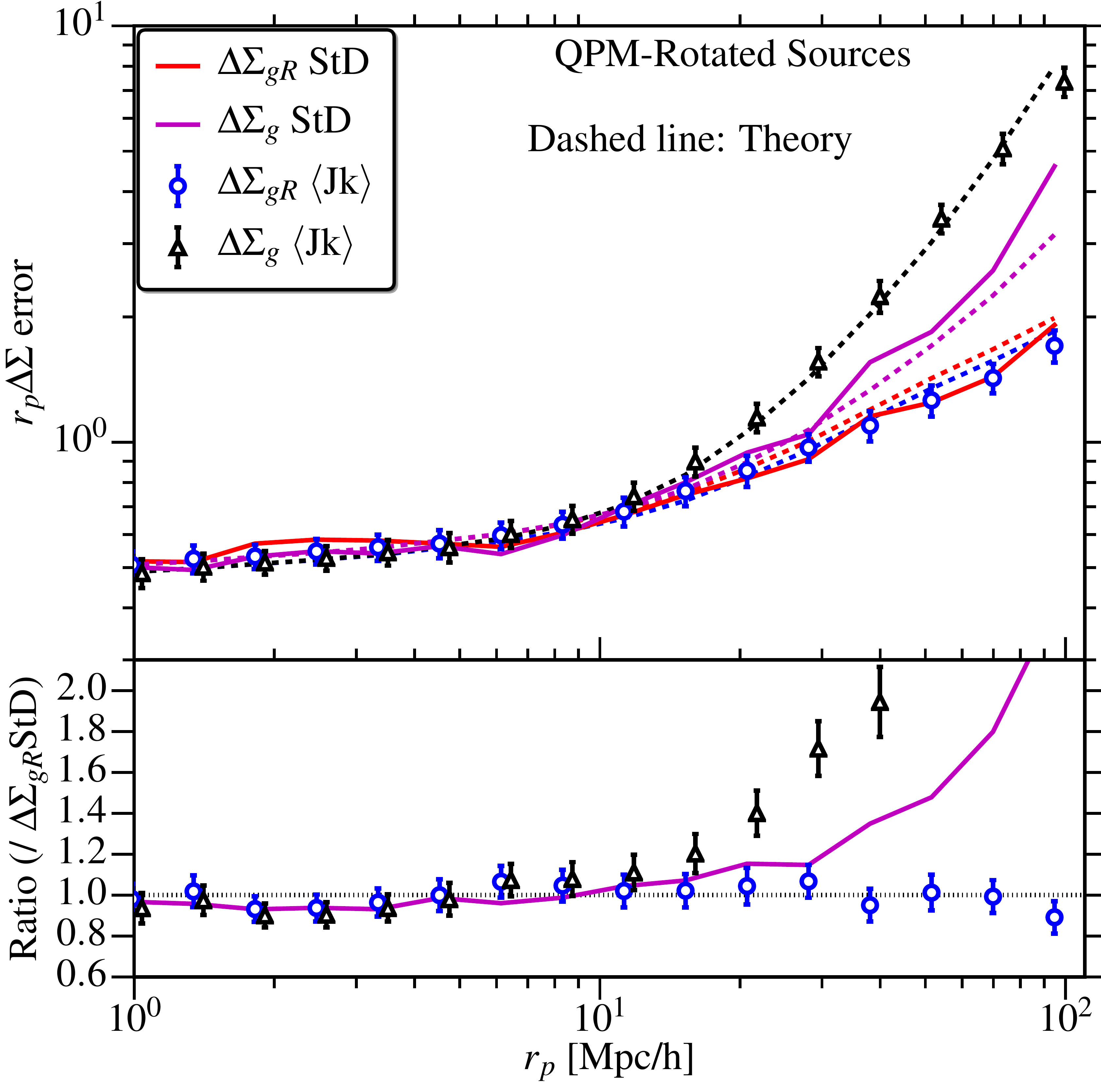}
		    \end{subfigure}

    	  \caption{ Same as Fig.~\ref{fig:SR_error}, using the QPM mocks as the lens sample with real sources (left 
	  		panel) and \referee{rotated (mock) sources} (right panel). 
			As in Fig.~\ref{fig:SR_error}, the errors with real sources are higher due to 
			contributions from systematics. The theoretical predictions are consistent except at large scales due to 
			systematics and the 
			effects of the LOWZ window function, which are not well captured by the idealized window function assumed in 
			the theoretical calculations. Also in the case of the real sources, the LOWZ jackknife
            errors are consistent with those for the QPM 
			mocks,
			suggesting that errors from the connected part of the covariance are subdominant.
    		}
	      \label{fig:qpm_error}
    	\end{figure*}

    		The left panel of Fig.~\ref{fig:qpm_error} shows the error estimates in the $\DS$
            measurements with the $\DS_{gR}$ estimator using  
			$N_R=10$, and with the $\DS_g$ estimator (without the measurement around random points
            subtracted). The jackknife 
    		errors with the $\DS_R$ subtraction are consistent with the error estimates using the standard deviation 
			across 200 QPM mocks. Also the jackknife errors for the LOWZ sample are consistent with the jackknife and 
    		standard deviation errors computed from the QPM mocks. This consistency confirms that the errors are 
    		dominated by the shape noise, $N_\gamma(P_{gg}+N_g)$ with some contributions from systematics, while 
			contributions  from cosmic
            variance and super-sample covariance (not included in the signal with the QPM mocks) are subdominant.
    		
			As in the case of the LOWZ lens sample, the theoretical predictions for the errors in the left panel of Fig.~\ref{fig:qpm_error} are 
			lower than the actual errors. This is due to the effect of systematics (there are no connected terms in this 
			case). In the right panel,  we show the 
			errors estimated by using the \referee{rotated (or mock) sources}. 
			In this case there are no shear correlations
            (either cosmological or due to systematics) and the theoretical predictions for the 
			errors are consistent with the measured errors in the case of $\DS_{gR}$. The
            discrepancies in the case of $\DS_g$ 
			are due to the idealized window function used for the extra term in the theory calculations.
		
%    		In Fig.~\ref{fig:qpm_corr} we also show the correlation matrices estimated using the jackknife and standard 
%    		deviation
%    		method for the QPM mocks and also compare it with the jackknife correlation matrix for the LOWZ sample. The 
%    		correlation matrices from standard deviation and jackknife are consistent (within estimation uncertainty 
%    		$\sim15\%$)
%    		and the LOWZ correlation matrix is also consistent with that for the QPM mocks, furthering reinforcing the 
%    		assertion that the errors are dominated by the shape noise.
    %      \begin{figure*}
    %         \centering
    %         \includegraphics[width=2\columnwidth]{qpm_DS_corr_comparison}
    %         \caption{ Comparison of jackknife correlation matrices for $\DS_g$ and $\DS_{gR}$ with the correlation 
    %         	matrix 
    %         	obtained using 200 QPM mocks, along with the $\DS_{gR}$ jackknife correlation matrix
    %            for the LOWZ sample.
    %         }
    %         \label{fig:qpm_corr}
    %      \end{figure*}      
    
   	\subsubsection{Randoms}\label{ssec:results_random_lens}
	   In this section, we measure $\DS$ by replacing the LOWZ galaxies with random lens catalogs. In
       this case, the covariance only has contributions from terms with lens shot noise, 
       $N_g(P_{\gamma\gamma}+N_\gamma)$, as there is no lens clustering, $P_{gg}=0$.  
	   We use 75 random samples that are the same 
	   size as the LOWZ sample, along with 10 additional random samples, which are used to compute $\DS_R$. In this 
	   section we only show results using the mock source sample, so $P_{\gamma\gamma}=0$.
		
\begin{figure}
	   		%\begin{subfigure}{\columnwidth}
	    	  \centering
		      \includegraphics[width=\columnwidth]{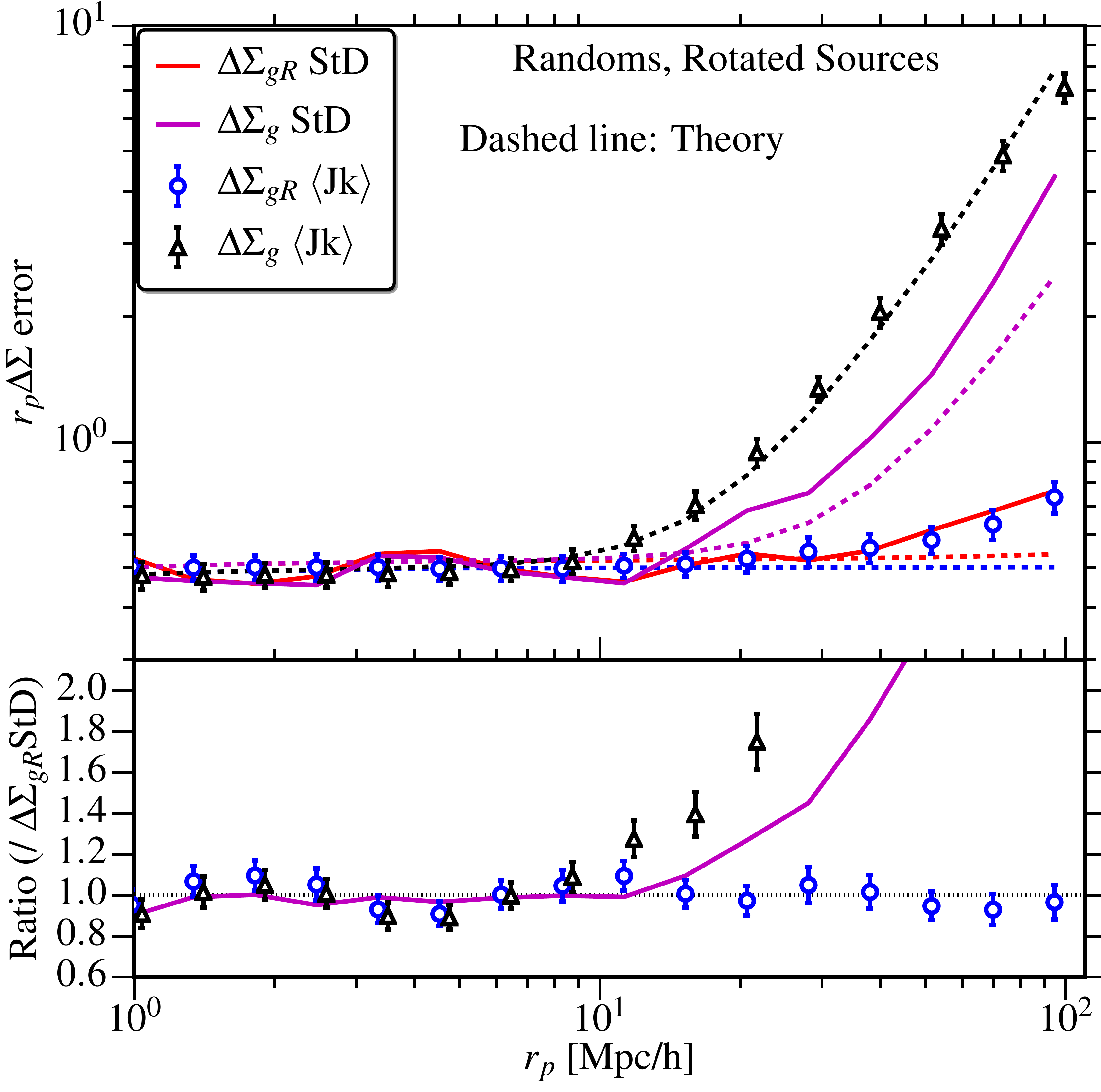}
		    %\end{subfigure}
%		   	\begin{subfigure}{\columnwidth}
%	    	  \centering
%		      \includegraphics[width=\columnwidth]{random_SR_DS_error}
%		    \end{subfigure}

    	  \caption{Same as Fig.~\ref{fig:SR_error}, now using 75 random realizations of the LOWZ lens
            catalog (no clustering) with \referee{rotated (mock) sources}. 
            Since the lens sample has no clustering and sources have no shear 
			correlations, the errors only include shot noise terms and hence scale as $1/r_p$,
            except at the largest scales 
			where the random points have some clustering as they match the selection function in LOWZ.
    		}
	      \label{fig:randoms_error}
		\end{figure}

		Fig.~\ref{fig:randoms_error} shows the errors in the \DS\ measurements using randoms lenses, with and 
		without $\DS_R$ subtracted out. Also shown are the error estimates using the standard deviation of the 
		signal measured across all 75 independent realizations. 
		 
		In the case of $\DS_{gR}$ errors from jackknife, the errors from the standard deviation and
        theory are consistent. The errors also 
		follow
		the expected $1/r_p$ scaling (no lens clustering in this case), except at the largest scales where there are 
		some deviations, possibly due to small amounts of large-scale power in the distribution of
        the random catalogs that enables them to match the selection 
		function of the LOWZ sample. Also, the 
		errors in the case of the random lenses are in  general lower than those for the LOWZ sample
        or the QPM mocks at large scales due to the effects of lens clustering.  Finally, for $\DS_{g}$ the errors do
        not follow the typical $1/r_p$ scaling because of the additional $W^2 N_\gamma$ term.

	\subsection{Putting it all together}	
		Using the results of the previous subsections, we can now understand the contributions of various terms in the 
		covariance, using both theoretical predictions and errors estimated using data and mocks.
		
\begin{figure*}
			\begin{subfigure}{\columnwidth}
		      \centering
		      \includegraphics[width=\columnwidth]{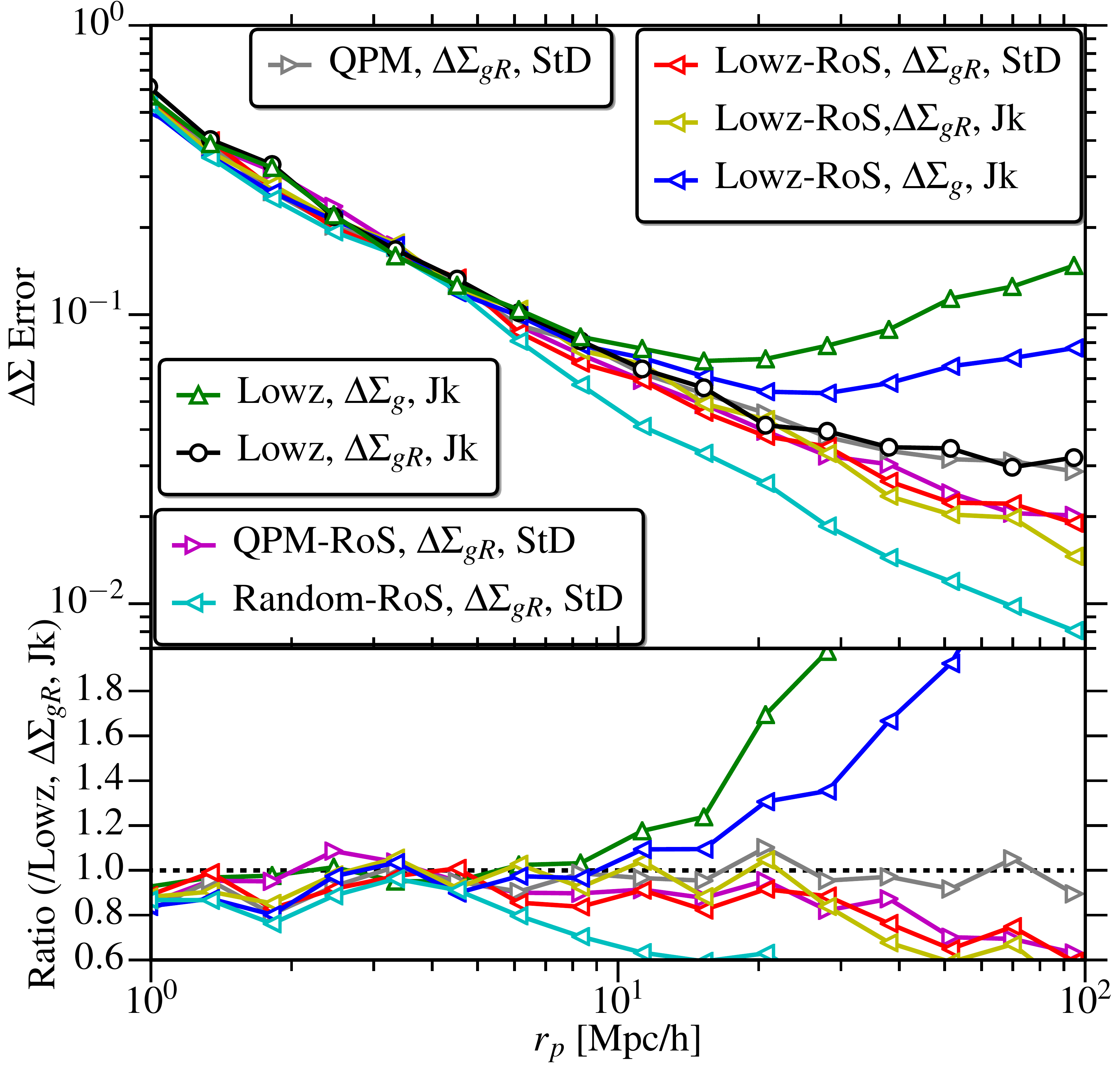}
		    \end{subfigure}
		    \begin{subfigure}{\columnwidth}
		      \centering
		      \includegraphics[width=\columnwidth]{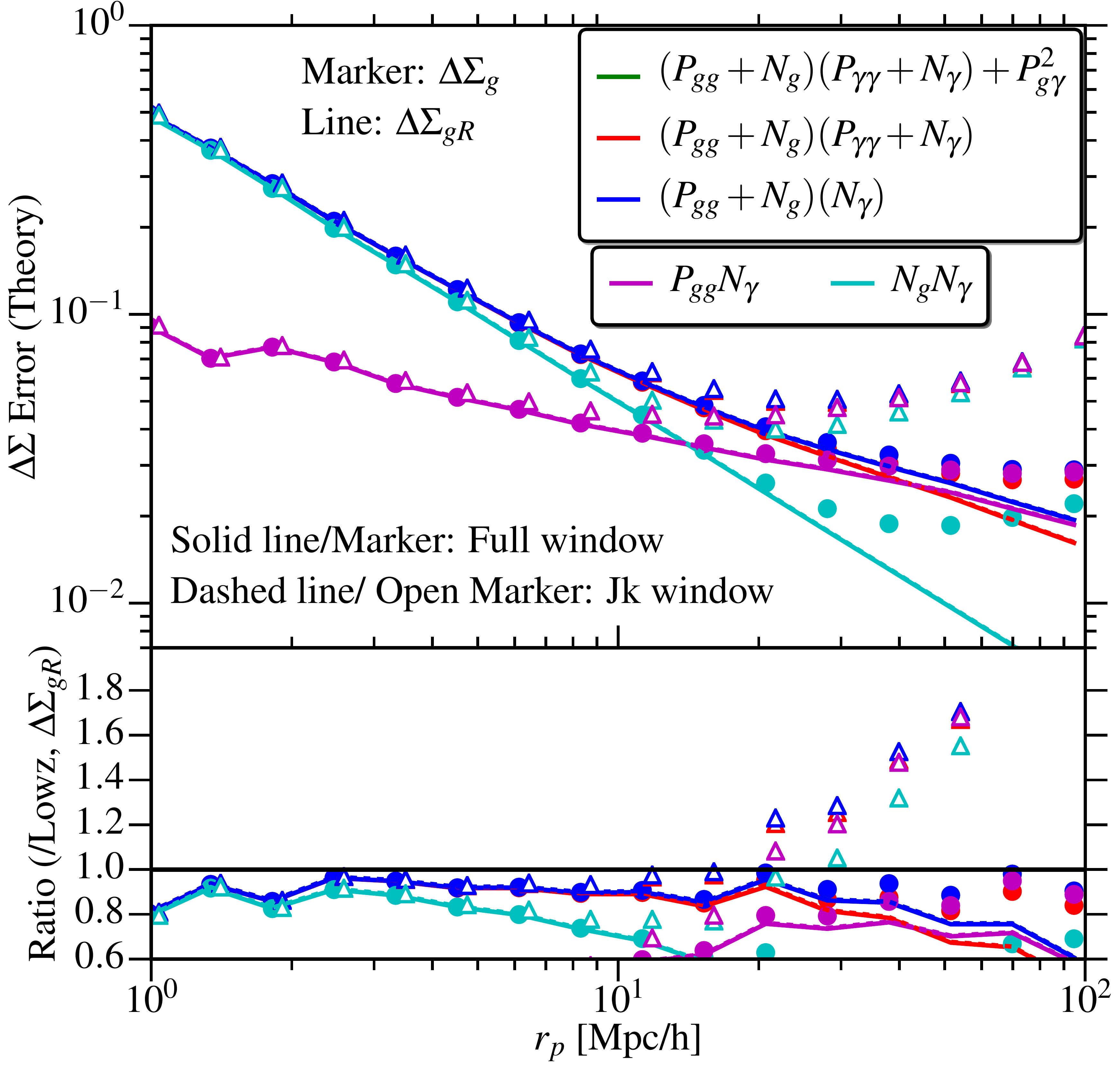}
		    \end{subfigure}
      		\caption{Figure summarizing the main error estimates discussed in this paper. Note that unlike in other 
				plots, 
				the y-axis here is the error on $\DS$ error without a factor of $r_p$. In the left
                panel, we show the errors estimated 
				using different combinations of the data and mock catalogs \referee{(RoS stands for ``Rotated 
				sources'' (or mock sources);} but
                unless explicitly mentioned, the curves use the 
				real sources). % (unless mentioned, errors use $\DS_{gR}$). 
				In the right panel, we show the error estimates from 
				different terms that contribute to the theoretical covariance using the notation of 
				Eq.~\eqref{eq:theory_GG_cov}. 
				In the bottom panels, the curves are divided by the jackknife errors on the real LOWZ $\DS_{gR}$, which 
				contains contributions from 
				systematics that are not included in the theoretical expressions, hence the ratios are systematically 
				below 1.
				For different curves, the power spectra terms not mentioned in the legend are set to
                zero, e.g.\ for the cyan 
				curves ($N_gN_\gamma$) all auto-power spectra are zero. There is still an $\{N_\gamma\}$ 
				term in this case in $\DS_{gR}$. %Similarly for other curves.
		    }
      		\label{fig:summary}
    	\end{figure*}

		In Fig.~\ref{fig:summary} we show the error estimates from various combinations of data and mocks (left 
		panel) and theory calculations using various terms in Eq.~\eqref{eq:theory_GG_cov} (right panel). In 
		Table~\ref{tab:summary} we also show various terms that contribute to various combinations of data and mocks.
		
		As shown in table~\ref{tab:summary}, we can use the random lenses with the \referee{rotated (mock) sources} 
		to compute the contributions of lens shot noise and source shape noise to the covariance. Including the
        lens samples with clustering 
		(LOWZ or QPM) then provides the contribution from the clustering of the lenses. Substituting
        the real sources with mock
		lenses (QPM) provides the contributions from shear correlations (systematics or cosmological). Thus we can study 
		all
		terms except for those arising from lens-source correlations ($P_{g\gamma}$ and $T_{g\gamma g\gamma}$) using the 
		mocks we have 
		used in this paper. Using more realistic simulations as in \cite{Shirasaki2016} will further
        allow a study of these 
		lens-source correlations terms, though as shown in previous sections, contributions from these terms are 
		subdominant when using the SDSS shape sample.  
		
		As demonstrated in Fig.~\ref{fig:summary}, at small scales the errors are dominated by the shot noise terms 
		$N_gN_\gamma$, where $N_g$ is the galaxy shot noise power spectrum and $N_\gamma$ is the shape noise power 
		spectrum. At larger scales ($r_p\gtrsim20\mpch$), the term involving the lens clustering, $P_{gg}N_\gamma$,
		starts dominating (in the literature this term is commonly referred to as ``correlated shape noise''). The 
		contributions
		from the shear power spectrum terms are in general small, with $P_{\gamma\gamma}$ and $P_{g\gamma}$ terms only 
		contributing $\sim10\%$ of the error even at $r_p\sim100\mpch$ for this particular survey. 
		In this work we did not compute the trispectrum terms ($T_{g\gamma g\gamma}$), but based on
        the comparison between the 
		LOWZ and QPM lens samples (see Fig.~\ref{fig:qpm_error}), we find no evidence that such terms are important 
		at any scale considered in this work for SDSS.
		
		We also show the contributions of the window function-dependent terms in $\DS_g$. As the size of
        the window function 
		increases, $W(k)$ approaches a delta function, $\delta_D(k)$.
		As a result, the contribution of these terms to the covariance decreases with increasing window 
		size, which is the reason why the jackknife errors have higher contributions than the standard deviation from the 
		mocks using full survey window. Once these terms are removed, the errors in $\DS_{gR}$ are consistent from
		both the jackknife and the full window. Normally we do expect $\DS_{gR}$ to be different
        between the jackknife and the full 
		window due to the edge effects (see results for clustering in appendix~\ref{appendix:clustering}). 
		However, in this work we only apply the jackknife to the lens sample while
		using the full source sample at all times. Hence the edge effects in the jackknife and full
        window cases are the same, though our theory curves under-predict these edge effects since we assumed an idealized window function 
		with circular symmetry and no holes (see Appendix~\ref{appendix:covariance}).

        \begin{table}
			\begin{tabular}{|c|c}
                \hline
                Lens-Shape  & Error Term \\ 
                Sample	& 			\\ \hline
                LOWZ-SDSS &     $(P_{gg}+N_g)(P_{\gamma\gamma}+N_\gamma)+P_{g\gamma}^2+T_{g\gamma g\gamma}$\\[5pt]
                &$+\{W^2(P_{\gamma\gamma}+N_\gamma)\}$\\ \hline
                QPM-SDSS &              $(P_{gg}+N_g)(P_{\gamma\gamma}+N_\gamma)+
                \{W^2(P_{\gamma\gamma}+N_\gamma)\}$
                          \\ \hline
                LOWZ-Mock &   $(P_{gg}+N_g)N_\gamma +\{W^2N_\gamma\}$\\ \hline
                QPM-Mock&              $(P_{gg}+N_g)N_\gamma +\{W^2N_\gamma\}$ \\ \hline
                % 
%                Randoms & SDSS & $\DS_g$ &                  $N_g(P_{\gamma\gamma}+N_\gamma)+
 %               \{W^2(P_{\gamma\gamma}+N_\gamma)\}$ \\ \hline
 %               Randoms & SDSS & $\DS_{gR}$ &               $N_g(P_{\gamma\gamma}+N_\gamma)$\\ \hline
                %
                Randoms-Mock&                  $N_gN_\gamma +\{W^2N_\gamma\}$\\ \hline
			\end{tabular}
			\caption{Table showing the main sources of statistical uncertainty for different
              combinations of lens, source and estimators 
				 in this work. We use the notation from Eq.~\eqref{eq:theory_GG_cov}, with
				$P_{gg}=b_gP_{\delta\delta}$, $N_g$ is the galaxy shot noise term, $N_\gamma$ is the shape noise and the 
				terms in curly brackets $\{\}$ involving $W^2$ are the window function-dependent
                covariance 
                contributions to $\DS_g$.
                }
			\label{tab:summary}
		\end{table}

   \subsection{Comparison of different error estimates}\label{ssec:results_subsample}
   		In this section we compare the error estimates from the jackknife method with 100 regions
        against those from taking the mean and 
		error on the mean from 100 subsamples. \referee{The primary motivation for this comparison is to test for edge effects and 
		to check whether the jackknife method underestimates the errors once the scales are close to
        the size of the subsamples.}		The subsamples were defined in the same way for both methods 
		%(Jackknife being leave one out) 
		and the division was only done on the lens sample. Each subsample/jackknife region is cross-
		correlated with the entire shape sample. Subsampling on the lens sample alone is sufficient in the shape noise-dominated regime, since the shape noise for different subsamples will be uncorrelated. In
        the case of other measurements, 
		e.g.\ clustering, the measurement across different subsamples will get correlated once the
        length scale approaches the size of 
		the subsample, and the errors will be underestimated in both cases.

   		\begin{figure}
		      \centering
		      \includegraphics[width=\columnwidth]{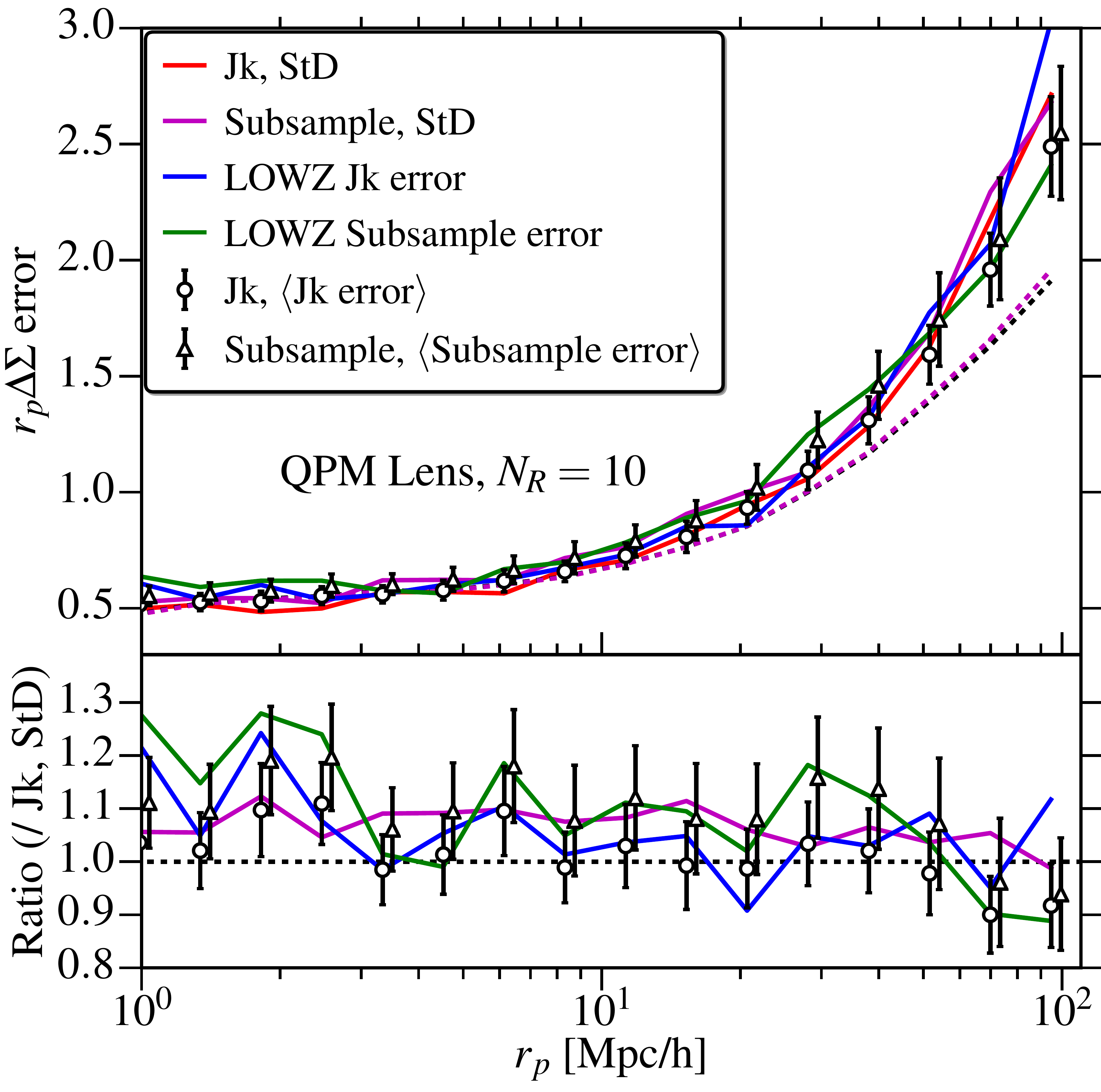}
      		\caption{Comparison of the error estimates obtained using the jackknife method and the standard deviation across 
				%the 
				100 subsamples of the lens samples (QPM except for the two lines labelled ``LOWZ''). The same subsamples 
				were used for both methods. All calculations use 
				$\DS_g-\DS_R$ with $N_R=10$. ``Jk, Std'' refer to the errors from the jackknife and
                from the standard deviation of the mock samples as defined in Sec.~\ref{subsec:covest}
				for $\DS_{gR}$ (signal, not the noise), while ``Subsample, Std'' refers to errors 
				from the standard deviations of $\DS_{gR}$ measured as the mean of subsamples in each realization. 
				\mean{\text{Jk error}} (\mean{\text{Subsample error}}) is the mean of the jackknife (subsample) error across 
				the realizations. Dashed lines are the corresponding theory predictions.
		    }
      		\label{fig:subsample_comparison}
    	\end{figure}
		
\begin{figure}
		      \centering
		      \includegraphics[width=\columnwidth]{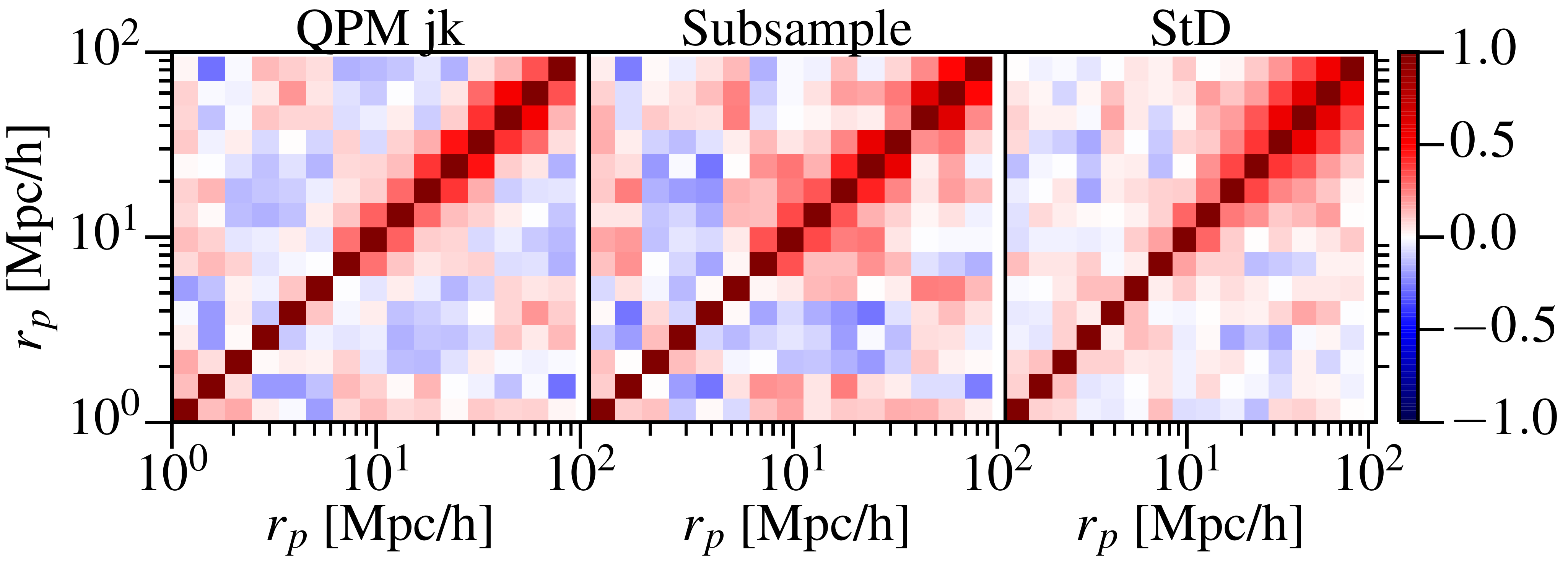}
      		\caption{Comparison of the correlation matrices for the QPM mocks from the jackknife, subsampling and standard deviation 
				methods across 200 realizations. All calculations use $\DS_g-\DS_R$ with $N_R=10$.
		    }
      		\label{fig:subsample_corr_comparison}
    	\end{figure}

		Fig.~\ref{fig:subsample_comparison} shows the comparison of three different error estimates: jackknife, 
		subsampling and
		standard deviation for the QPM mocks. All error estimates are consistent with each other (within the uncertainties), though the
		scatter in the subsampling errors is somewhat higher than the jackknife errors. In 
		Fig.~\ref{fig:subsample_corr_comparison} we also show the correlation matrix for the three different error 
		estimates; they are all consistent with each other.
	
\section{Conclusions}\label{sec:conclusion}

	In this work we have studied the behavior of covariances in galaxy-galaxy lensing measurements  
	using mock catalogs and theoretical predictions, including a comparison of two different
    galaxy-galaxy lensing estimators with different covariance properties. The mock catalogs 
    include randomly distributed lenses, QPM mocks that have similar clustering as LOWZ galaxies, and mock
    source catalogs obtained by randomly rotating the real SDSS source galaxies. Our main results 
    are summarized in Fig.~\ref{fig:summary} and Table~\ref{tab:summary}.

	For the SDSS, at small scales the 
    covariance is dominated by the pure uncorrelated shape noise, which is white, 
    but at larger scales, contributions from lens and shear correlations also 
    matter. Using mock source catalogs obtained by randomly rotating the sources, we show that the errors 
    are dominated by the terms involving the shape noise, $N_{\gamma}(P_{gg}+N_g)$.
    When using the real sources from SDSS, we also found evidence of contributions to the covariance from the 
    systematics to the covariance, even when subtracting random signal (i.e. using lens over-density). While our 
    theory calculations did not include the contributions from systematics or the connected term (including the 
    super-sample variance), the consistency of the covariances when using LOWZ and QPM mocks as lenses 
    demonstrates that contributions from the connected terms are subdominant and
    the differences between the theoretical predictions and the measurements in Fig.~\ref{fig:lowz_error} arise primarily 
    from the systematics.
    This conclusion in general depends on the survey configuration, and for different surveys
	the tradeoffs between shape noise and other covariance contributions must be re-evaluated.
           
    We also demonstrated that the additional variance seen on large scales when not subtracting the shear
    around random points is only partially contributed by systematics.  Even without systematics, on
    scales where correlated shape noise is important, using the sub-optimal estimator (lens density
    instead of over-density) 
for
    galaxy-galaxy lensing can reduce the per-bin $S/N$ by a substantial factor (up to a factor of 2
    on the largest scales considered, which are $\sim 10$ times the correlation length of the lens
    galaxies).
    Our covariance calculations suggest that this reduction in the covariance primarily arises from
    the removal of shear correlations (including shape noise), which only depends on the window function of the lens 
    sample.  Covariance estimation methods using subsamples
    (including the jackknife) have a smaller survey window and hence have a 
    higher contribution from this term compared to the
    full survey window. The tests using mock lens and mock source
    catalogs are consistent with this explanation.  Our results suggest that the discrepancy
    between the jackknife error estimates and standard deviation across different realizations observed
    by \cite{Shirasaki2016} (\referee{version 1}) can be explained by the fact that they used the lens density instead 
    of the lens over-density 
    in the galaxy-galaxy lensing measurements, given that the contributions from the additional covariance
    terms due to use of density depend on the survey window. \referee{This assertion has been confirmed in the updated
    work (version 2) of \cite{Shirasaki2016}.}

	In our calculations of covariances, we also identified the effects of the window function, which can be important 
	when 
	comparing the covariance estimations from empirical methods such as jackknife or subsampling, which divide the 
	survey 
	window into smaller parts. Since in our jackknife estimates we only split the lens sample, these effects are not 
	important in our lensing measurements. In appendix~\ref{appendix:clustering}, using clustering measurements, we show 
	that the window function effects can alter the covariance by up to 40-50\% on scales approaching
    the size of the subsamples. Finally, for both clustering (appendix~\ref{appendix:clustering}) and lensing 
    (section~\ref{ssec:results_subsample}), we also demonstrated using mocks that the jackknife errors are
    consistent with the errors from the subsampling methods for the scales that  are smaller than the subsample size 
    at the effective redshift of the sample.
%	We compared covariances for the mock lensing estimates, which do not include either cosmic variance or super-
%	sample covariance, with jackknife covariances from the real SDSS data.  Our results suggest that on the
%	scales used here, $\lesssim 100~h^{-1}$Mpc, 
%	there is no significant covariance contribution from
%	cosmic variance or super-sample covariance in the SDSS lensing measurements, which are dominated by
%	correlated shape noise. This conclusion depends on the survey configuration and for different surveys
%	the tradeoffs between correlated shape noise and other covariance contributions must be re-evaluated.
%	Some of the consistency checks demonstrated in this work may be useful, as are more
%	realistic mocks such as those in \cite{Shirasaki2016}.
	
%	Finally, we also compared the error estimates from jackknife with estimates from using the
%    standard deviation between the signals in the 100 regions used for the jackknife, finding them 
%	to be consistent. 
%This, however, is unlikely to be true when errors are not dominated by shape noise.\sukhdeep{In appendix B we show it to be more or less true even in absence of shape noise. This will not be true I think for scale greater than the size of subsample window, regime which we don't really explore.}
	
	Our results emphasize the importance of using the optimal galaxy-galaxy lensing estimator $\DS_g-\DS_R$
	even in the absence of systematics, for g-g lensing estimates that extend to scales above a few Mpc,
	to obtain better covariance properties and to enable use of internal error estimates like the
    jackknife. Our conclusions are also applicable to galaxy-CMB lensing cross correlations, as were done by 
    \cite{Singh2016}. Finally we recommend the use of the tests of covariances
	demonstrated in this work and those in \cite{Shirasaki2016} for ongoing and future surveys, to better understand 
	which terms are dominating the covariances.

\section*{Acknowledgements}

	We thank Fran\c{c}ois Lanusse, {Reiko Nakajima, Ben Wibking, Erin Sheldon} and Masahiro Takada for useful discussions related to 
	this work. We also thank Martin White and Jeremy 
	Tinker for providing the QPM mocks and the SDSS-I,II,III collaborations for providing the datasets used in this 
	work.
	
 SS acknowledges the support from John Peoples Jr.\ Presidential Fellowship from Carnegie Mellon University.
 RM acknowledges the support of the Department of Energy Early Career Award program.  US acknowledges support of NASA grant NNX15AL17G.

Funding for SDSS-III has been provided by the Alfred P. Sloan Foundation, the Participating Institutions, the National Science Foundation, and the U.S. Department of Energy Office of Science. The SDSS-III web site is http://www.sdss3.org/.

SDSS-III is managed by the Astrophysical Research Consortium for the Participating Institutions of the SDSS-III Collaboration including the University of Arizona, the Brazilian Participation Group, Brookhaven National Laboratory, Carnegie Mellon University, University of Florida, the French Participation Group, the German Participation Group, Harvard University, the Instituto de Astrofisica de Canarias, the Michigan State/Notre Dame/JINA Participation Group, Johns Hopkins University, Lawrence Berkeley National Laboratory, Max Planck Institute for Astrophysics, Max Planck Institute for Extraterrestrial Physics, New Mexico State University, New York University, Ohio State University, Pennsylvania State University, University of Portsmouth, Princeton University, the Spanish Participation Group, University of Tokyo, University of Utah, Vanderbilt University, University of Virginia, University of Washington, and Yale University.

        \bibliographystyle{mnras}
        \bibliography{papers,sukhdeep_cmb_paper,DS_error} %Please add new references in DS_error.bib

\onecolumn
\appendix
      \section{Covariance}\label{appendix:covariance}
      	\subsection{General case}\label{appendix:covariance_general}			
			Here we derive the expression for the covariance of the cross-correlation function of
            two fields with non-zero mean.   The results depend on the estimator used for that
            cross-correlation function, as we will show explicitly below (see also \citealt{Landy1993}), 
            and directly motivate the
            use of estimators that involve subtraction of the mean density for both fields.
			While we will use the example of clustering in this section, the results are in general true for any tracer 
			of large-scale structure. In Section~\ref{appendix:covariance_lensing} we will use the results from this 
			section to compute the covariance for the galaxy-shear cross-correlation function.
						
    		We are interested in the cross-correlation function of  two (biased) tracer fields, $g_X,g_Y$, 
			of the matter density field ($\rho_m$, not just $\delta\rho_m$)
			\begin{align}
		 		g_i&=(1+\delta_i+ n_i) M_i
		 	\end{align}
    		 where $i=X$ or $Y$, $ M_i$ is the mean value of the field (mean number density in the case of galaxies)
		 	 and $n_i$ is the noise in the tracer field (shot noise 
		 	 in the case of galaxies, shape noise in the case of shear). Hereafter, in this 
    		 section we will assume that the field is normalized so that $M_i=1$. 
		 	For notational compactness, we also define
			\begin{align}
				\hdelta_i&=\delta_i+n_i\\
				g_i&=1+\hdelta_i
			\end{align}
			In Fourier space
			\begin{align}
				\tilde g_i(\vec{k})&=\delta_D(\vec k)+\tilde\delta_i(\vec k)+\tilde n_i(\vec k)=\delta_D(\vec k)+\tilde
				\hdelta_i(\vec k),
			\end{align}
			where $\delta_D$ is the Dirac delta function.
			
    		 We can write the cross-correlation function of two fields as (analogous to normalized $\frac{DD}{RR}-1$)
			 \begin{align}
    		 	{\hxi}_{XY}(\vec r)= {\hxi}_{g_1g_2}(\vec r)=&\frac{1}{\mathcal V_W(\vec r)}\int d^3\vec r'
				W(\vec r'+\vec r)W(\vec r')
			\left[g_1(\vec r')g_2(\vec r'+\vec r)-1\right]
				\label{eq:correlation_def}\\
			 					=&\frac{1}{\mathcal V_W(\vec r)}\int d^3\vec r'W(\vec r'+\vec r)W(\vec r')
								\left[\hdelta_1(\vec r')\hdelta_2(\vec r'+\vec r)+
								\hdelta_1(\vec r')+
								\hdelta_2(\vec r'+\vec r)+1\right]-1
								%\frac{1}{\mathcal V_W(\vec r)}\int d^3\vec r'W(\vec r'+\vec r)W(\vec r')
								\\
							=&\frac{1}{\mathcal V_W(\vec r)}\int d^3\vec r'W(\vec r'+\vec r)W(\vec r')
							\hdelta_1(\vec r')\hdelta_2(\vec r'+\vec r)\\=&
							\frac{1}{\mathcal V_W(\vec r)}\int d^3\vec r'W(\vec r'+\vec r)W(\vec r')
							\left[\delta_1(\vec r')\delta_2(\vec r'+\vec r)+
							n_1(\vec r')n_2(\vec r'+\vec r)\right]\\
				{\hxi}_{XY}(\vec r)= {\hxi}_{g_1g_2}(\vec r)=&{\xi}_{XY}(\vec r)+{\xi}_{n_Xn_Y}(\vec r)
    		 \end{align} 
			where $g_1$ belongs to field $X$ and $g_2$ to field $Y$. $W(\vec r)$ is the survey window function.	
            We have assumed that the noise and $\delta_i$ have 
			zero mean and are also uncorrelated with each other on all scales. 
			The normalization factor is the integral over window functions
			\begin{align}
				\mathcal V_W (\vec r)=\int d^3\vec r' W(\vec r'+\vec r)W(\vec r')
				=\int\frac{ d^3\vec k}{(2\pi)^3}e^{-i\vec k\cdot\vec r}\tilde W(\vec k)\tilde W(-\vec k)
			\end{align}
			
			The covariance of the correlation function is given as
			\begin{align}
    		 	\text{Cov}(\widehat\xi_{g_1g_2}(\vec r_i)\widehat\xi_{g_3g_4}(\vec r_j))=&\mean{\hxi_{g_1g_2}(\vec r_i)
					\hxi_{g_3g_4}(\vec r_j)}-
				\mean{\hxi_{g_1g_2}(\vec r_i)}\mean{\hxi_{g_3g_4}(\vec r_j)},
				%\\[15pt]
			\end{align}
			where $g_1,g_3$ belong to field $X$ and $g_2,g_4$ belong to $Y$. Using Eq.~\eqref{eq:correlation_def} 
			\begin{align}
				\text{Cov}(\widehat\xi_{g_1g_2}(\vec r_i)\widehat\xi_{g_3g_4}(\vec r_j))=&
				\left\langle\frac{1}{\mathcal V_W(\vec r_i)\mathcal V_W(\vec r_j)}\int d^3 \vec r \int d^3 \vec r' 
				W(\vec r)W(\vec r')W(\vec r+\vec r_i)W(\vec r'+\vec r_j)\left[
				g_1(\vec r) g_2(\vec{r}+\vec{r}_i) g_3(\vec r')g_4(\vec{r}'+\vec{r}_j)-\right.\right.\nonumber\\&
				\left.\left.
				g_1(\vec{r})g_2(\vec{r}+\vec{r}_i)-
				g_3(\vec{r}')g_4(\vec{r}'+\vec{r}_j)+1
				\right]\vphantom{\frac{1}{\mathcal V_W(\vec r_i)\mathcal V_W(\vec r_j)}}\right\rangle%\nonumber\\&
				-\mean{\hxi_{g_1g_2}(\vec r_i)}\mean{\hxi_{g_3g_4}(\vec r_j)}
				\\
				\text{Cov}(\widehat\xi_{g_1g_2}(\vec r_i)\widehat\xi_{g_3g_4}(\vec r_j))=&
				\mean{\frac{1}{\mathcal V_W(\vec r_i)\mathcal V_W(\vec r_j)}\int d^3 \vec r \int d^3 \vec r'
				W(\vec r)W(\vec r')W(\vec r+\vec r_i)W(\vec r'+\vec r_j)
				g_1(\vec r) g_2(\vec{r}+\vec{r}_i) g_3(\vec r')g_4(\vec{r}'+\vec{r}_j)}-\nonumber\\&
				\frac{\mean{\hxi_{g_1g_2}(\vec r_i)}}{\mathcal V_W (\vec r_i)}-
				\frac{\mean{\hxi_{g_3g_4}(\vec r_j)}}{\mathcal V_W (\vec r_j)}-1
				-\mean{\hxi_{g_1g_2}(\vec r_i)}\mean{\hxi_{g_3g_4}(\vec r_j)}\label{eq:multi-term}
					%\\[15pt]
					%=&\mean{\hdelta_1\hdelta_2\hdelta_3\hdelta_4}+\mean{\hdelta_1\hdelta_3}+\mean{\hdelta_2\hdelta_4}-
					%\mean{\hdelta_{1}\hdelta_{2}}\mean{\hdelta_{3}\hdelta_{4}}
			\end{align}
			We use $\mean{g_1g_2g_3g_4}_{ij}$ as short-hand for the first term in
            Eq.~\eqref{eq:multi-term}, which we would like to simplify.
			\begin{align}
				\mean{g_1g_2g_3g_4}_{ij}=&\mean{\frac{1}{\mathcal V_W(\vec r_i)\mathcal V_W(\vec r_j)}\int d^3 \vec r 
				\int d^3 \vec r' 
				g_1(\vec{r}) g_2(\vec{r}+\vec{r}_i) g_3(\vec r')g_4(\vec{r}'+\vec{r}_j)
				W(\vec r)W(\vec r')W(\vec r+\vec r_i)W(\vec r'+\vec r_j)}%\\[15pt]
			\end{align}
			Writing the $g_i$ in terms of its Fourier space counterpart $\tilde g_i$, we get
			\begin{align}
				\mean{g_1g_2g_3g_4}_{ij}=&\frac{1}{\mathcal V_W(\vec r_i)\mathcal V_W(\vec r_j)}\int d^3 \vec r \int d^3 
				\vec r'
				\iiiint \prod_{n=1}^4\left[\frac{d^3 \vec k_n}{(2\pi)^3}\right]
				\iiiint\prod_{m=1}^4\left[\frac{d^3 \vec q_m}{(2\pi)^3}\tilde W(\vec q_m)\right]\nonumber\\&
				\times
				e^{i(\vec{k}_1-\vec{q}_1)\cdot\vec{r}}
				e^{i(\vec{k}_2-\vec{q}_2)\cdot(\vec{r}+\vec{r_i})}
				e^{i(\vec{k}_3-\vec{q}_3)\cdot\vec r'}e^{i(\vec{k}_4-\vec{q}_4)\cdot(\vec{r}'+\vec{r}_j)}
				\mean{\tilde g_1(\vec k_1)\tilde g_2(\vec k_2)
				\tilde g_3(\vec k_3)\tilde g_4(\vec k_4)}
				\\%[15pt]
				\mean{g_1g_2g_3g_4}_{ij}=&\frac{1}{\mathcal V_W(\vec r_i)\mathcal V_W(\vec r_j)}
				\iint \frac{d^3 \vec k_1}{(2\pi)^3}\frac{d^3 \vec k_3}{(2\pi)^3} 
				\iiiint\prod_{m=1}^4\left[\frac{d^3 \vec q_m}{(2\pi)^3}\tilde W(\vec q_m)\right]
				e^{-i(\vec{k}_1-\vec{q}_1)\cdot\vec{r}_i}e^{-i(\vec{k}_3-\vec{q}_3)\cdot\vec{r}_j}\nonumber\\&\times
				\mean{\tilde g_1(\vec k_1)\tilde 
				g_2(-\vec k_1+\vec q_1+\vec q_2)\tilde g_3(\vec k_3)\tilde g_4(-\vec k_3+\vec q_3+\vec q_4)}.
				\label{eq:g_fourier_space}
			\end{align}
			We have integrated over $d^3 \vec r$ and $d^3 \vec r'$ and then over $d^3 \vec k_2$ and $d^3 \vec k_4$, to 
			obtain the last expression.
			
			We now expand the four-point function into two separable parts: the connected or non-Gaussian component
			$\mean{\tilde \delta_1\tilde \delta_2\tilde \delta_3\tilde \delta_4}'$ and the Gaussian component, which 
			using Wick's theorem can be expanded as the sum of the product of two-point functions.
	\referee{
			\begin{align}
				\mean{g_1g_2g_3g_4}_{ij}
				=&\frac{1}{\mathcal V_W(\vec r_i)\mathcal V_W(\vec r_j)}
				\iint \frac{d^3 \vec k_1}{(2\pi)^3}\frac{d^3 \vec k_3}{(2\pi)^3} 
				\iiiint\prod_{m=1}^4\left[\frac{d^3 \vec q_m}{(2\pi)^3}\tilde W(\vec q_m)\right]
				e^{-i(\vec{k}_1-\vec{q}_1)\cdot\vec{r}_i}e^{-i(\vec{k}_3-\vec{q}_3)\cdot\vec{r}_j}
				\left[
				\mean{\tilde \delta_1\tilde \delta_2\tilde \delta_3\tilde \delta_4}'+\right.\nonumber\\&
				\mean{\tilde \delta_1\tilde \delta_2}\mean{\tilde \delta_3\tilde \delta_4}+
				\left[\mean{\delta_{D,1}\delta_{D,2}}+\mean{\tilde n_1\tilde n_2}\right]
				\mean{\tilde \delta_3\tilde \delta_4}+
				\mean{\tilde \delta_1\tilde \delta_2}\left[\mean{\delta_{D,3}\delta_{D,4}}+
				\mean{\tilde n_3\tilde n_4}\right]+
				\nonumber\\&
				\mean{\tilde \delta_1\tilde \delta_3}\mean{\tilde \delta_2\tilde \delta_4}+
				\left[\mean{\delta_{D,1}\delta_{D,3}}+\mean{\tilde n_1\tilde n_3}\right]
				\mean{\tilde \delta_2\tilde \delta_4}+
				\mean{\tilde \delta_1\tilde \delta_3}\left[\mean{\delta_{D,2}\delta_{D,4}}+
				\mean{\tilde n_2\tilde n_4}\right]+
				\nonumber\\&
				\mean{\tilde \delta_1\tilde \delta_4}\mean{\tilde \delta_2\tilde \delta_3}+
				\left[\mean{\delta_{D,1}\delta_{D,4}}+\mean{\tilde n_1\tilde n_4}\right]
				\mean{\tilde \delta_2\tilde \delta_3}+
				\mean{\tilde \delta_1\tilde \delta_4}\left[\mean{\delta_{D,2}\delta_{D,3}}+
				\mean{\tilde n_2\tilde n_3}\right]+
				\nonumber\\&
				\mean{\delta_{D,1}\delta_{D,2}{\tilde n_3\tilde n_4}}+\text{all perms}
				+
				\nonumber\\&
				\mean{\tilde n_1\tilde n_2\tilde n_3\tilde n_4}+\mean{\delta_{D,1}\delta_{D,2}\delta_{D,3}\delta_{D,4}}
				\left.\vphantom{\mean{\tilde \delta_1}}\right]%\vphantom to get braces size right. Argument is same as 
				%used by first left
				\label{eq:general_cov_long}
			\end{align}
		}
			We have omitted the positional arguments for $\tilde \delta_i$ and $\tilde n_i$, which
            are the same as for 
			$g_i$ in Eq.~\eqref{eq:g_fourier_space}. We defined $\delta_{D,i}=\delta_D(\vec k_i)$ and 
			$\mean{\tilde\delta_i\tilde\delta_j}= P_{ij}(\vec k_i)\delta_D(\vec k_i+\vec 
			k_j)$, where $P_{ij}(\vec k)$ is the power spectrum. 
			
			Simplifying, the terms involving $\mean{\xi_{12}(\vec{r}_i)}\mean{\xi_{34}(\vec{r}_j)}$
			cancel out, and using the fact 
			that $g_1$ and $g_3$ belonged to field $X$ and $g_2$ and $g_4$ belonged to field $Y$, we can write the 
			covariance as
			\begin{align}
				\text{Cov}=&\left[\frac{\mathcal V_{W}(\vec r_i-\vec r_j)}{\mathcal V_W(\vec r_i)\mathcal V_W(\vec r_j)}
				\int \frac{d^3 \vec k}{(2\pi)^3} e^{-i\vec{k}\cdot
				\vec{r}_i}e^{i\vec{k}\cdot\vec{r}_j}
				\widehat P_{XX}(\vec k)\widehat P_{YY}(\vec k)+
				\frac{\mathcal V_{W}(\vec r_i+\vec r_j)}{\mathcal V_W(\vec r_i)\mathcal V_W(\vec r_j)}
				\int \frac{d^3 \vec k}{(2\pi)^3} e^{-i
				\vec{k}\cdot\vec{r}_i}e^{-i\vec{k}\cdot
				\vec{r}_j}\widehat P_{XY}(\vec k)\widehat P_{XY}(\vec k)+T_{XYXY}\right]\nonumber\\&
				+\left\{\frac{1}{\mathcal V_W(\vec r_i)\mathcal V_W(\vec r_j)}
				\int \frac{d^3 \vec k}{(2\pi)^3}e^{-i\vec{k}\cdot\vec{r}_i}e^{i\vec{k}\cdot
				\vec{r}_j}\tilde 
				W(\vec k)\tilde W(-\vec k) 
				\left(\widehat P_{YY}(\vec k)+\widehat P_{XX}(\vec k)\right)\right\}\nonumber\\&
				+\left\{\frac{1}{\mathcal V_W(\vec r_i)\mathcal V_W(\vec r_j)}
				\int \frac{d^3 \vec k}{(2\pi)^3}e^{-i\vec{k}\cdot\vec{r}_i}e^{-i\vec{k}\cdot
				\vec{r}_j}\tilde 
				W(\vec k)\tilde W(\vec k) 
				\left(\widehat P_{XY}(\vec k)+\widehat P_{XY}(\vec k)\right)\right\}
				\label{eq:general_cov_full}
			\end{align}

			Here $\widehat P_{ij}=P_{ij}+P_{ij,N}$, where $P_{ij,N}$ is the noise power spectrum. $T_{XYXY}$ is the 
			connected term. % and $\mathcal V_W$ is the physical volume of the survey.
			To simplify expressions, we have assumed that the power spectrum is a slowly varying function of $k$ 
			and that we are working with modes much smaller than  the survey 
			size, so that $P(\vec k-\vec q)\approx P(\vec k)$ and then $P(\vec k)$ can be moved out
            of the window function integrals. For scales much smaller than the survey size,
            $\mathcal V_W(\vec r)\rightarrow V_W$, \referee{where $V_W$ is the physical volume of the survey}, 
            the expression simplifies to the more familiar form
            
            \begin{align}
				\text{Cov}=&\left[\frac{1}{V_W}
				\left(\int \frac{d^3 \vec k}{(2\pi)^3} e^{-i\vec{k}\cdot
				\vec{r}_i}e^{i\vec{k}\cdot\vec{r}_j}
				\widehat P_{XX}(\vec k)\widehat P_{YY}(\vec k)+\int \frac{d^3 \vec k}{(2\pi)^3} e^{-i
				\vec{k}\cdot
				\vec{r}_i}e^{-i\vec{k}\cdot
				\vec{r}_j}\widehat P_{XY}(\vec k)\widehat P_{XY}(\vec k)\right)+T_{XYXY}\right]\nonumber\\&
				+\left\{\frac{1}{V_W^2}
				\int \frac{d^3 \vec k}{(2\pi)^3}e^{-i\vec{k}\cdot\vec{r}_i}e^{i\vec{k}\cdot
				\vec{r}_j}\tilde 
				W(\vec k)\tilde W(-\vec k) 
				\left(\widehat P_{YY}(\vec k)+\widehat P_{XX}(\vec k)\right)\right\}\nonumber\\&
				+\left\{\frac{1}{V_W^2}
				\int \frac{d^3 \vec k}{(2\pi)^3}e^{-i\vec{k}\cdot\vec{r}_i}e^{-i\vec{k}\cdot
				\vec{r}_j}\tilde 
				W(\vec k)\tilde W(\vec k) 
				\left(\widehat P_{XY}(\vec k)+\widehat P_{XY}(\vec k)\right)\right\}\nonumber\\&
				\label{eq:general_cov_short}
			\end{align}
			The terms in square brackets ([ ]) are the usual covariance terms
            while the terms in braces 
			(\{\}) arise when the means of the fields are not subtracted. These additional contributions depend on 
			the survey window function and become less important as the survey size increases. In the case of a
            large uniform 
			survey, $\lim_{V_W\rightarrow\infty}\tilde{W}(\vec k)=\delta_D(\vec k)$.  
			As a result, the terms in braces (\{\}) approach zero faster (under the assumption $\widehat P(\vec k=0)=0$) 
			and the two estimators (correlating mean zero field or correlating mean non-zero fields) are equivalent. 
			However, in case $\widehat P(\vec k=0)\neq0$, e.g., due to shot noise in case of galaxies, the (\{\}) terms approach 
			the value of $P({\vec k=0})$. We also emphasize that this additional contribution to the
            covariance will 
			be present in the analysis in Fourier space as well.

		\subsection{Projected Case}\label{appendix:covariance_projected}
%\color{referee_C}
%\referee{ %latex doesnot like this
			The projected correlation function is defined as the integral of the 3-d correlation function over the line of 
			sight separation, $\Pi$.
			\begin{align}
				\widehat w(r_p)=\int_{\Pi_\text{min}}^{\Pi_\text{max}} d\Pi W(\Pi) \hxi(r_p,\Pi)
			\end{align}
			where $W(\Pi)$ is the line-of-sight weight function (not necessarily the same as the window function).
			  
			To compute the covariance, we start with Eq.~\eqref{eq:general_cov_long}, carry out the line-of-sight 
			integrals assuming the integration length is long (getting delta functions of the form $\delta_D(\vec 
			k_{\parallel,i}-\vec q_{\parallel,i})$). Then carrying out integrals involving the line-of-sight window 
			functions, we assume that the relevant line-of-sight modes are small, such that power spectrum is 
			only dependent on the projected modes ($k_\parallel\ll k_\perp$, $ P(\vec k)\approx P(\vec k_{\perp})$)
			
%			\begin{align}
%				\text{Cov}=&\int d\Pi_i\int d\Pi_j W(\Pi_i)W(\Pi_j)
%				\left[\frac{\mathcal V_{W}(\vec r_i-\vec r_j)}{\mathcal V_W(\vec r_i)\mathcal V_W(\vec r_j)}
%				\int \frac{d^3 \vec k}{(2\pi)^3} e^{-i\vec{k}\cdot
%				\vec{r}_i}e^{i\vec{k}\cdot\vec{r}_j}
%				\widehat P_{XX}(\vec k_{\perp})\widehat P_{YY}(\vec k_{\perp})\right.\nonumber\\&
%				+\left.
%				\frac{\mathcal V_{W}(\vec r_i+\vec r_j)}{\mathcal V_W(\vec r_i)\mathcal V_W(\vec r_j)}
%				\int \frac{d^3 \vec k}{(2\pi)^3} e^{-i
%				\vec{k}\cdot\vec{r}_i}e^{-i\vec{k}\cdot
%				\vec{r}_j}\widehat P_{XY}(\vec k_{\perp})\widehat P_{XY}(\vec k_{\perp})+T_{XYXY}\right]\nonumber\\&
%				%
%				+\int d\Pi_i\int d\Pi_jW(\Pi_i)W(\Pi_j)
%				\left\{\frac{1}{\mathcal V_W(\vec r_i)\mathcal V_W(\vec r_j)}
%				\int \frac{d^3 \vec k}{(2\pi)^3}e^{-i\vec{k}\cdot\vec{r}_i}e^{i\vec{k}\cdot
%				\vec{r}_j}\tilde 
%				W(\vec k_\perp)\tilde W(-\vec k_\perp) L_W^2
%				\left(\widehat P_{YY}(\vec k_{\perp})+\widehat P_{XX}(\vec k_{\perp})\right)\right.\nonumber\\&
%				%
%				+\left.
%				\frac{1}{\mathcal V_W(\vec r_i)\mathcal V_W(\vec r_j)}
%				\int \frac{d^3 \vec k}{(2\pi)^3}e^{-i\vec{k}\cdot\vec{r}_i}e^{-i\vec{k}\cdot
%				\vec{r}_j}\tilde 
%				W(\vec k_\perp)\tilde W(\vec k_\perp)L_W^2 
%				\left(\widehat P_{XY}(\vec k_{\perp})+\widehat P_{XY}(\vec k_{\perp})\right)\right\}
%				\label{eq:w_cov}
%			\end{align}
%			Where $\vec r_i=\vec r_{p,i}+\Pi_i$. 
%			Integrating over $k_\parallel$ leads to
			\begin{align}
				\text{Cov}=&\int d\Pi W_Y(\Pi)W_Y(\Pi)%\int d\Pi_j W(\Pi_j)
				\left[\frac{\mathcal V_{W}(\vec r_i-\vec r_j)}{\mathcal V_W(\vec r_i)\mathcal V_W(\vec r_j)}
				\int \frac{d^2 \vec k_\perp}{(2\pi)^2} e^{-i\vec{k_\perp}\cdot
				\vec{r}_{p,i}}e^{i\vec{k_\perp}\cdot\vec{r}_{p,j}}
				\widehat P_{XX}(\vec k_{\perp})\widehat P_{YY}(\vec k_{\perp})
				\right.\nonumber\\&
				+\left.
				\frac{\mathcal V_{W}(\vec r_i+\vec r_j)}{\mathcal V_W(\vec r_i)\mathcal V_W(\vec r_j)}
				\int \frac{d^2 \vec k_\perp}{(2\pi)^2} 
				e^{-i\vec{k_\perp}\cdot\vec{r}_{p,i}}e^{-i\vec{k_\perp}\cdot\vec{r}_{p,j}}
				\widehat P_{XY}(\vec k_{\perp})\widehat P_{XY}(\vec k_{\perp})
				+T_{XYXY}\right]\nonumber\\&
				+\int d\Pi W_Y(\Pi)W_Y(\Pi)
				\left\{\frac{L_W}{\mathcal V_W(\vec r_i)\mathcal V_W(\vec r_j)}
				\int \frac{d^2 \vec k_\perp}{(2\pi)^2}e^{-i\vec{k_\perp}\cdot\vec{r}_{p,i}}e^{i\vec{k_\perp}\cdot
				\vec{r}_{p,j}}\tilde 
				W_X(\vec k_\perp)\tilde W_X(-\vec k_\perp) 
				\left(\widehat P_{YY}(\vec k_{\perp})+\widehat P_{XX}(\vec k_{\perp})\right)%\delta_D(\Pi_j-\Pi_i)
				\right.\nonumber\\&
				+\left.
				\frac{L_W}{\mathcal V_W(\vec r_i)\mathcal V_W(\vec r_j)}
				\int \frac{d^2 \vec k_\perp}{(2\pi)^2}e^{-i\vec{k_\perp}\cdot\vec{r}_{p,i}}e^{-i\vec{k_\perp}\cdot
				\vec{r}_{p,j}}\tilde 
				W_X(\vec k_\perp)\tilde W_X(\vec k_\perp) 
				\left(\widehat P_{XY}(\vec k_{\perp})+\widehat P_{XY}(\vec k_{\perp})%\delta_D(\Pi_j+\Pi_i)
				\right)
				\right\}
				\label{eq:w_cov_full0}
			\end{align}
% To get the \{\} terms, we need to go dig deeper into full integrals. That is because there is integral $d^3q$ over window function along with $e^{iq.r}$ term. In case of only $r_\perp$, that integral give los window length as well.
%
			Here we distinguished between the window functions of tracers X and Y, $L_W$ is the line-of-sight length of 
			the window function (of $X$) and we ignore the edge effects along the line-of-sight. 
			Thus the volume element can be written as 
			\begin{equation}
				\mathcal{V}_W(\vec r_p)=\mathcal A_W(\vec r_p)L_W
			\end{equation}
			$\mathcal A_W$ is the physical survey area at the lens redshift.
			
			Note that $\widehat P_{YY}$ can in principle be evaluated at a 
			different epoch as $Y_1$ and $Y_2$ 
			are at separation $\Pi_i$ and $\Pi_j$, i.e., $P_{YY}(k_{\perp})\sim P_{YY}(k_{\perp}\frac{\chi_z}{\chi_z+
			\Pi})$ where $\chi_z$ is the line-of-sight distance to the mean redshift where we are evaluating the covariance.
			Under the assumption that the power spectrum evolution within the $\Pi_\text{max}$ limits is small, 
			we keep $P_{YY} (k_\perp)$ (ignoring its $\Pi$ dependence), and simplify the expression as 
			\begin{align}
				\text{Cov}=&
				\left[\frac{\mathcal A_{W}(\vec r_{p,i}-\vec r_{p,j})}{\mathcal A_W(\vec r_{p,i})\mathcal A_W(\vec 
				r_{p,j})}
				\frac{\Delta\Pi_2}{L_W}
				\int \frac{d^2 \vec k_\perp}{(2\pi)^2} e^{-i\vec{k_\perp}\cdot
				\vec{r}_{p,i}}e^{i\vec{k_\perp}\cdot\vec{r}_{p,j}}
				\widehat P_{XX}(\vec k_{\perp})\widehat P_{YY}(\vec k_{\perp})
				\right.\nonumber\\&\left.
				+\frac{\mathcal A_{W}(\vec r_i+\vec r_j)}{\mathcal A_W(\vec r_i)\mathcal A_W(\vec r_j)}
				\frac{\Delta\Pi_2}{L_W}
				\int \frac{d^2 \vec k_\perp}{(2\pi)^2} 
				e^{-i\vec{k_\perp}\cdot\vec{r}_{p,i}}e^{-i\vec{k_\perp}\cdot\vec{r}_{p,j}}
				\widehat P_{XY}(\vec k_{\perp})\widehat P_{XY}(\vec k_{\perp})+T_{XYXY}
				\right]\nonumber\\&
				+%\int d\Pi_i\int d\Pi_jW(\Pi_i)W(\Pi_j)
				\frac{\Delta\Pi_2 L_W^2}{\mathcal V_W(\vec r_{p,i})\mathcal V_W(\vec r_{p,j})}
				\left\{
				\int \frac{d^2 \vec k_\perp}{(2\pi)^3}e^{-i\vec{k_\perp}\cdot\vec{r}_{p,i}}e^{i\vec{k_\perp}\cdot
				\vec{r}_{p,j}}\tilde 
				W(\vec k_\perp)\tilde W(-\vec k_\perp) 
				\left(\widehat P_{YY}(\vec k_{\perp})+\widehat P_{XX}(\vec k_{\perp})\right)\right.\nonumber\\&
				+\left.
				\int \frac{d^2 \vec k_\perp}{(2\pi)^3}e^{-i\vec{k_\perp}\cdot\vec{r}_{p,i}}e^{-i\vec{k_\perp}\cdot
				\vec{r}_{p,j}}\tilde 
				W(\vec k_\perp)\tilde W(\vec k_\perp) 
				\left(\widehat P_{XY}(\vec k_{\perp})+\widehat P_{XY}(\vec k_{\perp})\right)\right\}
				\label{eq:w_cov_full}
			\end{align}
			where we defined
			\begin{align}
				\Delta\Pi_2&=\int d\Pi W(\Pi)W(\Pi)\\
				\Delta\Pi_1&=\int d\Pi W(\Pi)
			\end{align}
			For the case of galaxy clustering, we assume $W(\Pi)$ is a top-hat function for $\Pi\in[-100,100]$, which leads to
			$\Delta\Pi_2=\Delta\Pi_1=200\mpch$. 
%}		
\color{black}
		\subsection{Galaxy lensing case}\label{appendix:covariance_lensing}
	\newcommand{\hgamma}{\ensuremath{\widehat{\gamma}}}
			
			We now use the formalism of Appendix~\ref{appendix:covariance_general} and~\ref{appendix:covariance}
			to derive the
            covariance for the galaxy-galaxy lensing case. We will assume the same sky coverage for
            the lens and shape samples. Note that the shear is a mean-zero field since lensing is
            only sensitive to the matter density contrast, and hence some of the terms in  
			Eq.~\eqref{eq:general_cov_full} will drop out.
	
			We begin by defining the observed shear as the sum of the true shear and noise.
			\begin{align}
				\hgamma=\gamma+\gamma_N
			\end{align}			
			We also assume that the mean shear around random points is not subtracted.  In that case, 
			%for lens galaxies in a thin redshift slice $z_l$, 
			the galaxy-shear (projected) cross-correlation can be 
			written as
			\begin{align}
				\Delta\Sigma(\vec r_p)=&\frac{1}{\mathcal V_W(\vec r)}\int d^3\vec r \Sigma_c(z_l,z_s)
				g(\vec r)\hgamma(\vec r+\vec r_p)
				W_\gamma(\vec r+\vec r_p)W_g(\vec r)\\
				=&\frac{1}{\mathcal V_W(\vec r)}
				\int d^3\vec r \Sigma_c(z_l,z_s)
				(1+\hdelta)(\vec r)\hgamma(\vec r+\vec r_p)W_\gamma(\vec r+\vec r_p)W_g(\vec r)\\
				%=&\frac{1}{A_W}\int d^2\vec r \hdelta(\vec r)\hgamma(\vec r+\vec r_p)+\int d^2\vec r\hgamma(\vec r+
				%\vec r_p)\\
				\Delta\Sigma(\vec r_p)= &\frac{1}{\mathcal V_W(\vec r)}
			\int d^3\vec r \Sigma_c(z_l,z_s)\hdelta(\vec r)\hgamma(\vec r+\vec r_p)W_\gamma(\vec r+\vec r_p)W_g(\vec r)
			\end{align}
%			Since we are interested in projected correlation function, the integral over line-of-sight $r_\parallel$ can 
%			be written using $r_\parallel\sim \frac{c}{H(z)}dz$ and we get 
%			\begin{align}
%				w_{g\gamma}(\vec r_p)= &\frac{1}{\int dz p(z)\frac{c}{H(z)}\int d^2\vec rW_\gamma(\vec r+\vec 
%				r_p)W_g(\vec r)}
%				\int dz p(z) \frac{c}{H(z)}\int d^2\vec r \hdelta(\vec r)\hgamma(\vec r+\vec r_p)W_\gamma(\vec r+\vec 
%				r_p)W_g(\vec r)\\
%				=&
%			\end{align}
			
			%All distances are measured at the lens redshift.
			The covariance of two $\Delta\Sigma$ is			
			\begin{align}
				\text{Cov}(\Delta\Sigma_{g_1\gamma_2}(\vec r_{p,i})\Delta\Sigma_{g_3\gamma_4}(\vec r_{p,j}))=&
				\mean{\Delta\Sigma_{g_1\gamma_2}(\vec r_{p,i})\Delta\Sigma_{g_3\gamma_4}(\vec r_{p,j})}-
				\mean{\Delta\Sigma_{g_1\gamma_2}(\vec r_{p,i})}\mean{\Delta\Sigma_{g_3\gamma_4}(\vec r_{p,j})}
			\end{align}
			Following the derivation in appendix~\ref{appendix:covariance_general} and~\ref{appendix:covariance} 
			and noting that the
            shear has a mean of zero, 
			and $\Delta\Sigma$ is a projected galaxy-matter correlation function, the full covariance analogous to 
			Eq.~\eqref{eq:w_cov_full0} is
%			\begin{align}\label{eq:GG_cov}
%				\text{Cov}=&\left[\frac{A_W}{\mathcal A_W(\vec r_{p_i})\mathcal A_W(\vec r_{p_j})}
%				\left(\int\frac{ d^2\vec k}{(2\pi)^2} e^{-i\vec{k}\cdot\vec{r_p}_i}
%				e^{i\vec{k}\cdot\vec{r_p}_j}
%				\widehat P_{gg}(\vec k)\widehat P_{\gamma\gamma}(\vec k)+\int \frac{ d^2\vec k}{(2\pi)^2} 
%				e^{-i\vec{k}\cdot\vec{r_p}_i}e^{-i\vec{k}\cdot
%				\vec{r_p}_j}\widehat P_{g\gamma}(\vec k)\widehat P_{g\gamma}(\vec k)\right)+T_{ij}
%				\right]\nonumber\\&
%				+\left\{\frac{2\Pi_\text{max}}{\mathcal A_W(\vec r_{p_i})\mathcal A_W(\vec r_{p_j})}
%				\int \frac{ d^2\vec k}{(2\pi)^2}e^{-i\vec{k}\cdot\vec{r_p}_i}e^{i\vec{k}\cdot
%				\vec{r_p}_j}\tilde W(\vec k)\tilde W(-\vec k) \widehat P_{\gamma\gamma}(\vec k)\right\}
%			\end{align}
	%\color{referee_C}		
			\begin{align}
				\text{Cov}=&\left[\int d\Pi W(\Pi)W(\Pi)
				\frac{\mathcal V_{W}(\vec r_i-\vec r_j)}{\mathcal V_W(\vec r_i)\mathcal V_W(\vec r_j)}
				\int \frac{d^2 \vec k_\perp}{(2\pi)^2} e^{-i\vec{k_\perp}\cdot
				\vec{r}_i}e^{i\vec{k_\perp}\cdot\vec{r}_j}
				\widehat P_{gg}(\vec k_{\perp})\widehat P_{\delta\delta}(\vec k_{\perp})\right]
				\nonumber\\&
				+\left[\int d\Pi W(\Pi)W(\Pi)
				\frac{\mathcal V_{W}(\vec r_i+\vec r_j)}{\mathcal V_W(\vec r_i)\mathcal V_W(\vec r_j)}
				\int \frac{d^2 \vec k_\perp}{(2\pi)^2} 
				e^{-i\vec{k_\perp}\cdot\vec{r}_{p,i}}e^{-i\vec{k_\perp}\cdot\vec{r}_{p,j}}
				\widehat P_{g\delta}(\vec k_{\perp})\widehat P_{g\delta}(\vec k_{\perp})+T_{g\gamma g\gamma}
				\right]\nonumber\\&
				+\int d\Pi W(\Pi)W(\Pi)
				\left\{\frac{L_W^2}{\mathcal V_W(\vec r_i)\mathcal V_W(\vec r_j)}
				\int \frac{d^2 \vec k_\perp}{(2\pi)^3}e^{-i\vec{k_\perp}\cdot\vec{r}_i}e^{i\vec{k_\perp}\cdot
				\vec{r}_j}\tilde 
				W(\vec k_\perp)\tilde W(-\vec k_\perp) 
				\widehat P_{\delta\delta}(\vec k_{\perp})\right\}
				\label{eq:GG_cov_full}
			\end{align}
			where the lensing window function is 
			\begin{align}
			W(\Pi)=\overline \rho\frac{\Sigma_c(\chi_s,\chi_l)}{\Sigma_c(\chi_s,\chi_l+\Pi)}
			\end{align} 
			The line-of-sight integral in the terms involving $P_{\delta\delta}$ leads to 
			the shear auto-correlation function, and the final expression is 
			\begin{align}
				\text{Cov}=&\left[%\int d\Pi_i W(\Pi_i)W(\Pi_i)
				\frac{\mathcal V_{W}(\vec r_{p,i}-\vec r_{p,j})}{\mathcal V_W(\vec r_{p,i})\mathcal V_W(\vec r_{p,j})}
				\int \frac{d^2 \vec k_\perp}{(2\pi)^2} e^{-i\vec{k_\perp}\cdot
				\vec{r}_i}e^{i\vec{k_\perp}\cdot\vec{r}_{p,j}}\cos{2\phi_{k,i}}\cos{2\phi_{k,j}}
				\widehat P_{gg}(\vec k_{\perp})\Sigma_c^2\left(\frac{\sigma_\gamma^2}{n_s}+P_{\kappa\kappa}\right)
				\right]
				\nonumber\\&
				+\left[%\int d\Pi_i W(\Pi_i)W(-\Pi_i)
				\frac{\mathcal V_{W}(\vec r_{p,i}+\vec r_{p,j})\Delta\Pi_2}{\mathcal V_W(\vec r_{p,i})\mathcal V_W(\vec 
				r_{p,j})}
				\int \frac{d^2 \vec k_\perp}{(2\pi)^2} 
				e^{-i\vec{k_\perp}\cdot\vec{r}_{p,i}}e^{-i\vec{k_\perp}\cdot\vec{r}_{p,j}}
				\cos{2\phi_{k,i}}\cos{2\phi_{k,j}}\overline\rho^2
				\widehat P_{g\delta}(\vec k_{\perp})\widehat P_{g\delta}(\vec k_{\perp})+T_{g\gamma g\gamma}
				\right]\nonumber\\&
				+
				\left\{\frac{L_W^2}{\mathcal V_W(\vec r_{p,i})\mathcal V_W(\vec r_{p,j})}
				\int \frac{d^2 \vec k_\perp}{(2\pi)^3}e^{-i\vec{k_\perp}\cdot\vec{r}_{p,i}}e^{i\vec{k_\perp}\cdot
				\vec{r}_{p,j}}\cos{2\phi_{k,i}}\cos{2\phi_{k,j}}
				\tilde W(\vec k_\perp)\tilde W(-\vec k_\perp) 
				 \Sigma_c^2\left(\frac{\sigma_\gamma^2}{n_s}+P_{\kappa\kappa}\right)\right\}
				\label{eq:GG_cov_full}
			\end{align}
			where the convergence power spectrum is
			\begin{align} 
				P_{\kappa\kappa}(k)=\int_0^{\chi_s} d\chi \frac{\overline \rho}{\Sigma_c(\chi,\chi_s)}\frac{\overline 
				\rho}{\Sigma_c(\chi,\chi_s)}P_{\delta\delta}\left(k\frac{\chi_l}{\chi}
				\right)\label{eq:convergence}
			\end{align}
			For lensing, using the full lens and source redshift distribution, we compute $\Delta\Pi_1\approx900\mpch$ and $
			\Delta\Pi_2\approx700\mpch$.
	\color{black}
			%\referee{Here $k$ is the projected mode, $k\equiv k_\perp$,} 
			Again as in appendix~\ref{appendix:covariance_general}, the terms in square brackets ([ ]) are the usual 
			covariance terms for the mean zero fields and the terms in curly brackets (\{ \}) are
			additional contribution from terms involving $\delta_{D,i}$, arising from the
            sub-optimal estimator without the mean subtracted.
			
			Note that this additional contribution only depends on the window function of the lens sample and is 
			independent of the clustering or number density of the lens sample. Hence, this noise
            term is consistent across the real lens galaxy sample and uniformly-distributed random points,
            which is why the subtraction of the shear around random points removes this contribution to the
			covariance. Also the window function-dependence of this term is the reason why the jackknife/subsample 
			methods of estimating errors shows increased contribution from this term compared to the
            standard deviation across independent mock catalogs, since the effective window function
            for the subsamples is 
			smaller.

	\subsection{Numerical estimates}
		\subsubsection{Clustering}
			To compute numerical estimates, we assume angular symmetry for both the power spectra
            and the window function.
			Further, in the case of galaxy clustering, $X\equiv Y$ and $P_{XX}=b_g^2P_{\delta\delta}$.
			After carrying out the angular and line-of-sight integrals (for the projected correlation function) 
			in Eq.~\eqref{eq:w_cov_full}, we get
			\begin{align}
				\text{Cov}(w_{gg})=&\left[2 \frac{\mathcal A_W(|\vec r_{p,i}-\vec r_{p,j}|)}
				{\mathcal A_W( r_{p,i})\mathcal A_W( r_{p,j})}\frac{\Delta\Pi_2}{L_W}
				\int \frac{d k k}{2\pi} J_0(kr_{p,i})J_0(kr_{p,j})
				\left(b_g^2P_{\delta\delta}(k)+\frac{1}{n_g}\right)^2+T_{gggg}\right]\nonumber\\&
				+4\left\{\frac{2\Delta\Pi_{\text{2}}}{\mathcal A_W(r_{p,i})\mathcal A_W(r_{p,j})}
				\int \frac{d k k}{2\pi}J_0(kr_{p,i})J_0(kr_{p,j})
				\tilde 
				W(k)\tilde W(k) 
				\left( b_g^2 P_{\delta\delta}( k)+\frac{1}{n_g}\right)\right\}
				\label{eq:numerical_clustering_cov_full}
			\end{align}
			where $J_n$ is the bessel function of order $n$, $\Pi_\text{max}$ is the line-of-sight integration length 
			and
			\begin{align}\label{eq:AW_numerical}
				\mathcal A_W (r)=
				\int\frac{ dk \, k}{2\pi}J_0(kr)\left(W(k)\right)^2
			\end{align}
			\begin{align}
				\mathcal A_W (|\vec r_i-\vec r_j|)=
				\int\frac{ dk \, k}{2\pi}J_0(kr_i)J_0(kr_j)\left(W(k)\right)^2.
			\end{align}
			For the window function, we assume a circular geometry with a survey area of $9000$
            degree$^2$, with a mean 
			redshift of $z=0.27$. $W(k)$ is defined as
			\begin{align}
%				W(\vec k)=&\int d^2r e^{i\vec k\cdot \vec r} W(\vec r)
%						=2\pi\int dr r J_0(kr)\\%did angular integral.. assumed W(r)=1
%						=&2\pi R^2 \frac{J_1(kR)}{kR}\\
				W(k)&= 2\pi R^2\frac{J_1(kR)}{kR} % 
			\end{align}
			where $R\approx1275\mpch$ is the physical scale corresponding to 95~degrees at $z=0.27$.
			
			Finally, to get the covariance for bins in $r_p$, $\text{COV}_\text{bin}$, we integrate the covariance in 
			Eq.~\eqref{eq:numerical_clustering_cov_full} as 
			\begin{align}\label{eq:cov_bin}
				\text{COV}_\text{bin}=\frac{\int_{r_{p,i,l}}^{r_{p,i,h}}dr_{p,i}'r_{p,i}'
										\int_{r_{p,j,l}}^{r_{p,j,h}}dr_{p,j}'r_{p,j}' \text{COV}(r_{p,i}',r_{p,j}')}
						{\int_{r_{p,i,l}}^{r_{p,i,h}}dr_{p,i}'r_{p,i}'\int_{r_{p,j,l}}^{r_{p,j,h}}dr_{p,j}'r_{p,j}'}
			\end{align}
			where $r_{p,i,l},r_{p,i,h}$ are the lower and upper limits of the bins, respectively.

		\subsubsection{Galaxy Lensing}
		%\color{referee_C}
			We carry out the angular integrals in Eq.~\eqref{eq:GG_cov_full}, to get
			\begin{align}
				\text{Cov}(\DS)=&\left[\frac{\mathcal A_W(\vec r_{p,i}-\vec r_{p,j})}{\mathcal A_W(\vec r_{p,i})
				\mathcal A_W(\vec r_{p,j})}\frac{1}{L_W}
				\int \frac{d k k}{2\pi} J_2(kr_{p,i})J_2(kr_{p,j})\Sigma_c^2
				\left(b_g^2P_{\delta\delta}(k)+\frac{1}{n_g}\right)
				\left(P_{\kappa\kappa}(k)+\frac{\sigma_{\gamma}^2}{n_s}\right)
				\right]\nonumber\\&
				+\left[\frac{\mathcal A_W(\vec r_{p,i}-\vec r_{p,J})}{\mathcal A_W(\vec r_{p,i})\mathcal 
				A_W(\vec r_{p,J})}\frac{\Delta\Pi_2}{L_W}
				\int \frac{d k k}{2\pi} J_2(kr_{p,i})J_2(kr_{p,j})
				\left(b_gr_{cc}\overline\rho P_{\delta\delta}(k)\right)^2
				+T_{g\gamma g\gamma}\right]\nonumber\\&
				+\left\{\frac{1}{\mathcal A_W(\vec r_{p,i})\mathcal A_W(\vec r_{p,j})}
				\int \frac{d k k}{2\pi}J_2(kr_{p,i})J_2(kr_{p,j})
				\tilde 
				W(k)\tilde W(k) \Sigma_c^2
				\left(P_{\kappa\kappa}(k)+\frac{\sigma_{\gamma}^2}{n_s}\right)
				\right\}%\nonumber\\&
				\label{eq:numerical_DS_cov_full}
			\end{align}
		\color{black}
			\text{Cov}(\DS) is then integrated to get the covariance in bins as described in Eq.~\eqref{eq:cov_bin}.
			
	\section{Clustering results}\label{appendix:clustering}
		In this appendix we present the comparison of different estimators and error estimation methods for the
		galaxy clustering measurements.
		
		We begin by defining the standard Landy-Szalay (LS) estimator for clustering \citep{Landy1993} 
		\begin{equation}
				\widehat{\xi}_{LS}(r_p,\Pi)=\frac{(D-R)^2}{RR}=\frac{DD-2DR+RR}{RR},
				\label{LSxi}
		\end{equation}
		where $\xi$ is the three dimensional correlation function, 
		$r_p$ is the projected separation on the sky, 
		and $\Pi$ is the line-of-sight separation between the pair of galaxies. The use of 
		$D-R$ indicates that we are correlating over-density fields. The estimator can then be expanded into its 
		standard pair counting form. $DD$ denotes summation over all galaxy-galaxy 
		pairs within the bin, $DR$ are the cross pairs between galaxies and randoms, and $RR$ are the random-random pairs.

		In addition, we define the basic estimator from pair counting
		\begin{equation}
				\widehat{\xi}_{s2}(r_p,\Pi)=\frac{DD}{RR}-1
				\label{xi0}
		\end{equation}
		
		Motivated by the galaxy-galaxy lensing estimator without subtraction of the mean galaxy
        density, we also define the estimator correlating an 
        over-density field ($D-R
		$) with a density field ($D$)
		\begin{equation}
				\widehat{\xi}_{s1}(r_p,\Pi)=\frac{D(D-R)}{RR}=\frac{DD-DR}{RR}
				\label{xi_GG}
		\end{equation}

		We  want to work with projected correlation functions, analogous to galaxy-galaxy lensing. Thus we integrate 
		\xigg\ over the line-of-sight to obtain the projected correlation function \wgg.
		\begin{equation}
			\widehat{w}_{gg}(r_p)=\int^{\Pi_{\text{max}}}_{-\Pi_{\text{max}}}\widehat{\xi}(r_p, \Pi)\mathrm{d}\Pi.
		\end{equation}
		The choice of $\Pi_\text{max}$ depends on two considerations: we want to choose large $\Pi_\text{max}$ to 
		capture the full correlation function and to mitigate the effect of redshift space distortions 
		\citep{Kaiser1987}. However, in a survey of finite redshift window, 
        the bins at large $\Pi$ are also noisier which
		increases the noise in the projected correlation function as well. In this work we use $\Pi_\text{max}=100\mpch$ 
		with linear bins of size $d\Pi=10\mpch$. 
		
		\begin{figure*}
			\begin{subfigure}{0.48\columnwidth}
		    	  \centering
	    		  \includegraphics[width=\columnwidth]{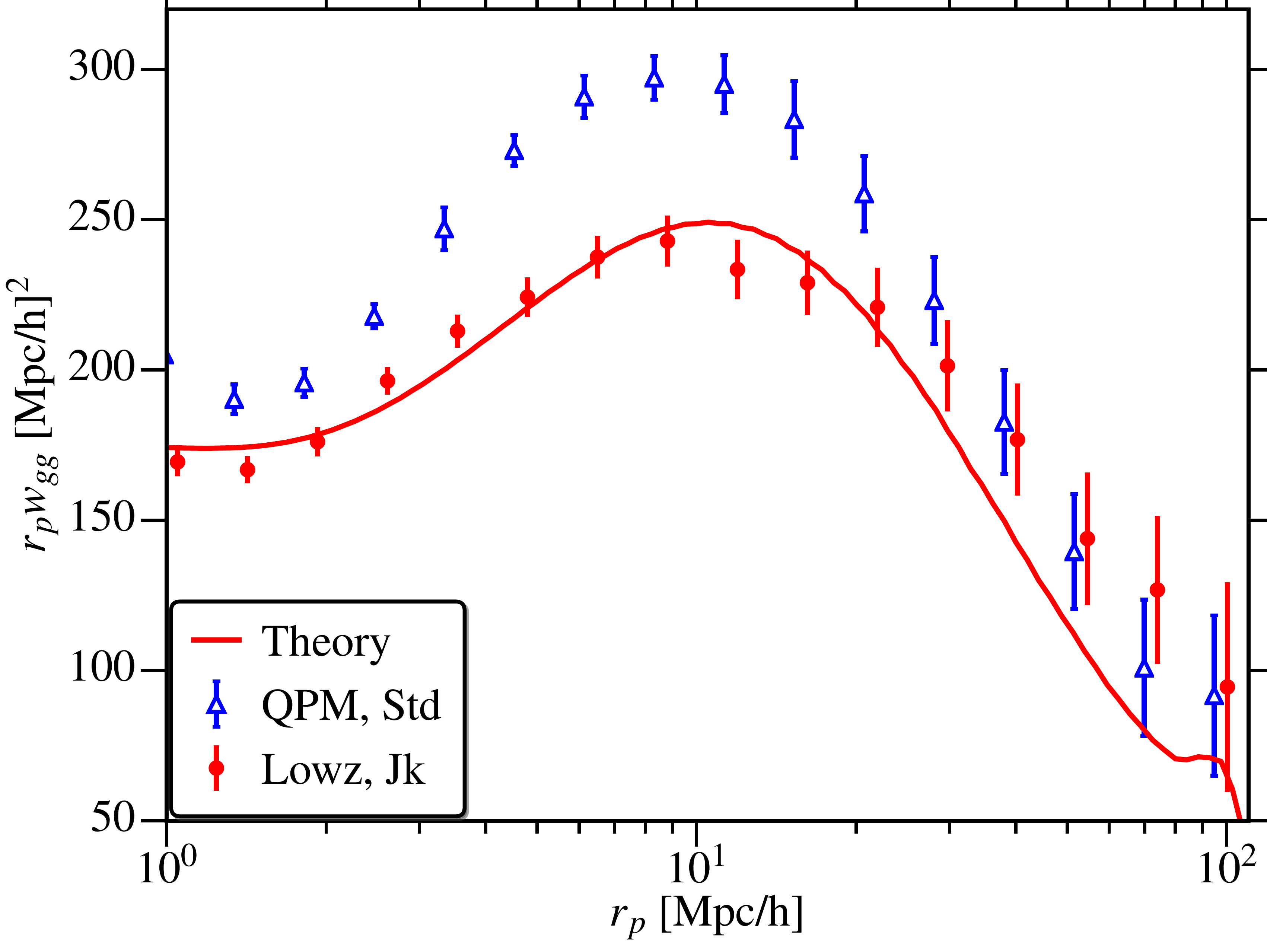}
			      \label{fig:qpm_wgg}
			      \caption{}
	    	\end{subfigure}
			\begin{subfigure}{0.48\columnwidth}
		    	  \centering
	    		  \includegraphics[width=\columnwidth]{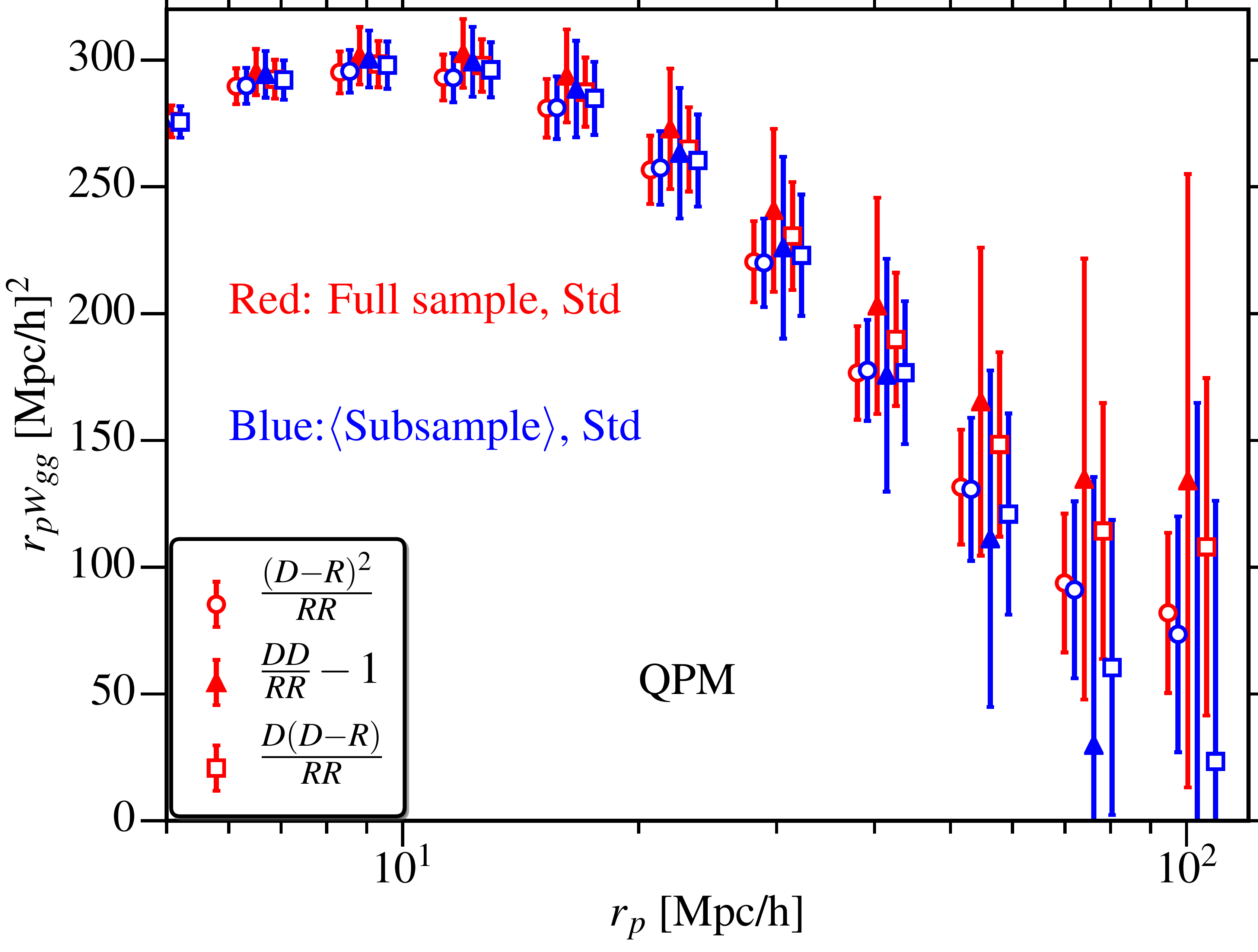}
			      \caption{}
	    	\end{subfigure}
	       \caption{a) Comparison of the projected clustering for the LOWZ sample and the QPM
             mocks. The LOWZ errors 
	  			are from the jackknife method, while the QPM values are the mean and standard deviation across 150 
				realizations. The red line is the Planck 2015 \lcdm\ prediction along with the best-fitting bias from 
				\citet{Singh2016}; note that the points on large scales have correlated errors.
				b) Comparison of clustering measurements using different estimators (note different $r_p$ range). 
				$\mean{\text{Subsample}}$ refers 
				to the mean signal across the subsamples, in each realization. We then take the mean and standard 
				deviation of $\mean{\text{Subsample}}$ across realizations.
	  		}
			\label{fig:qpm_wgg}
    	\end{figure*}
	
		In Fig.~\ref{fig:qpm_wgg} we show the clustering measurement for the LOWZ sample as well as for one realization 
		of the QPM mocks, with jackknife errors for both. At small scales, the clustering between
        the mocks and data does not 
		agree very well, with a maximum difference of order $\sim30\%$. This is expected, since the QPM mocks
        are generated using low-resolution simulations, which only resolve the large-scale density
        field \citep{White2014}. Here we compare the  
		estimators and error estimations self-consistently from the QPM mocks, and thus the failure to
        exactly match the LOWZ sample clustering is not important.
	
		\begin{figure*}
			\begin{subfigure}{0.48\columnwidth}
		   	  \centering
	    	  \includegraphics[width=\columnwidth]{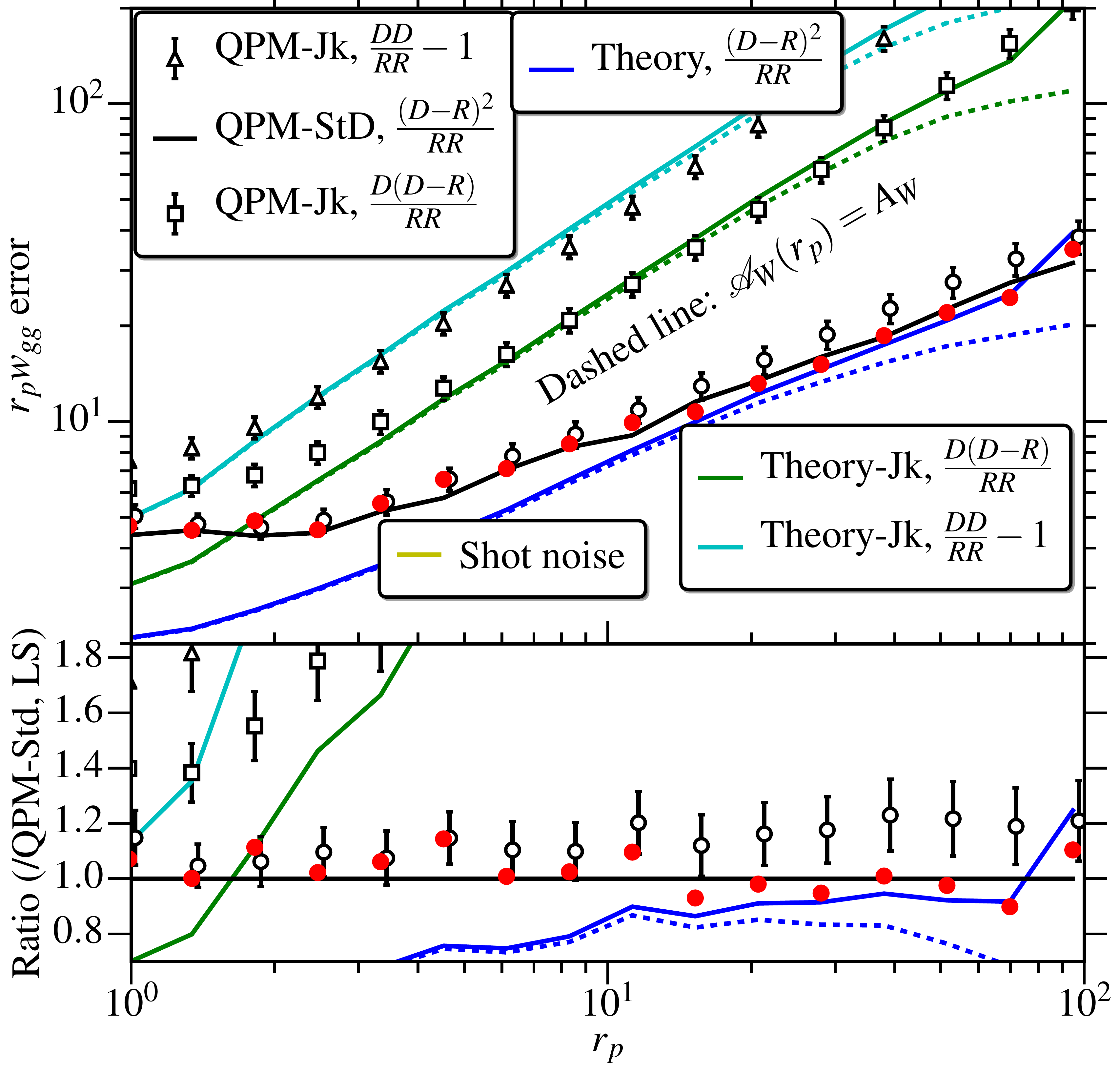}%{qpm_wgg_err}
		  	\end{subfigure}
			\begin{subfigure}{0.48\columnwidth}
		   	  \centering
	    	  \includegraphics[width=\columnwidth]{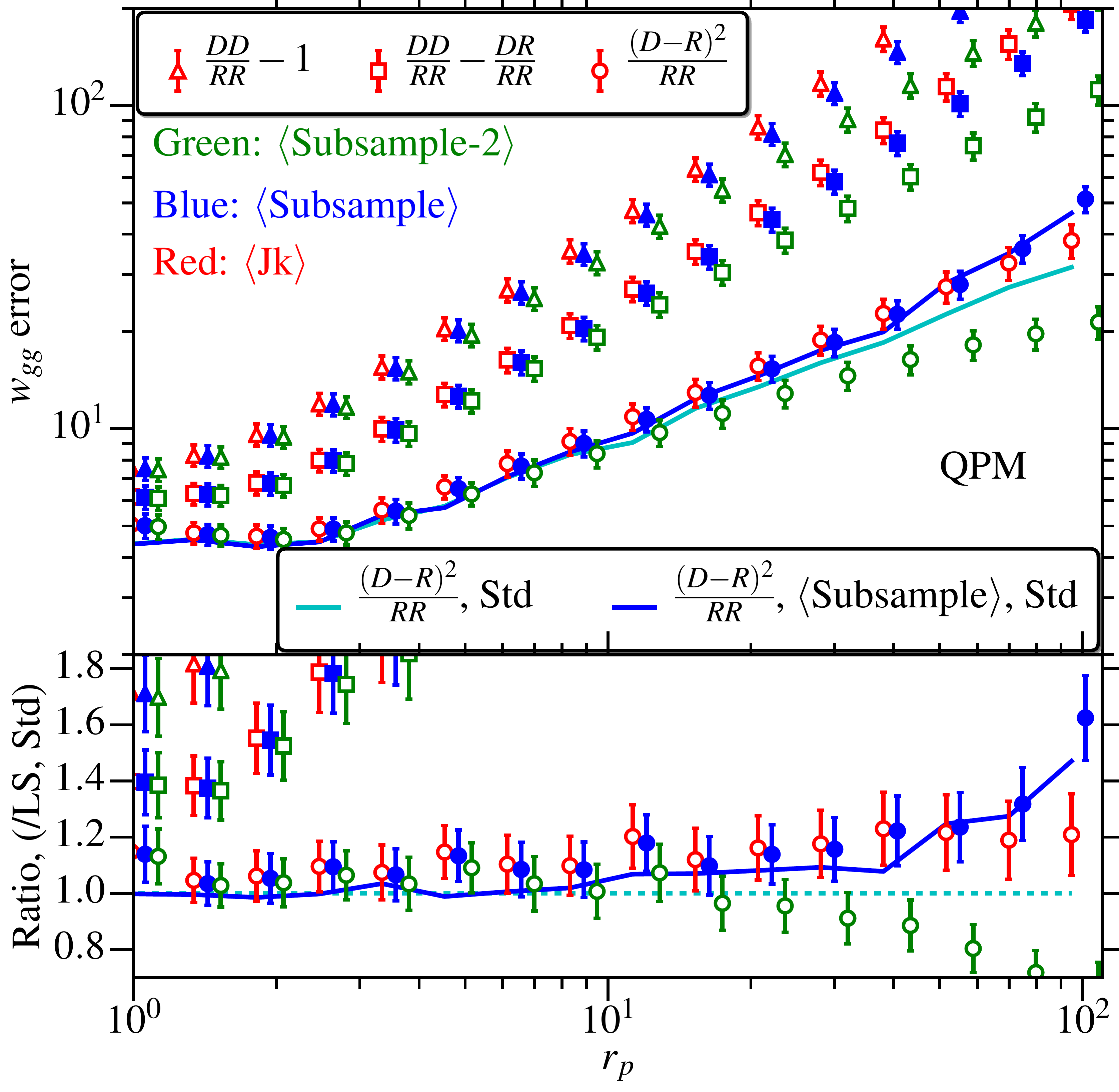}%{randoms_wgg_err}
		  	\end{subfigure}
    	  \caption{\emph{Left panel}: Comparison of errors for different clustering estimators defined in 
	  		Eqs.~\eqref{LSxi}--\eqref{xi_GG}.  
	  		The Landy-Szalay ($\frac{(D-R)^2}{RR}$) estimator, Eq.~\eqref{LSxi}, 
			gives the lowest error followed by the estimator in  
			Eq.~\eqref{xi_GG} and then the estimator in Eq.~\eqref{xi0}.
 	  		Also shown are the jackknife errors for the LS estimator, 
			with the jackknife overestimating the errors by $\sim10-20\%$ at all scales. We additionally show the 
			estimated  
			errors from the theory predictions. Note that the theory estimates use the linear theory$+$halofit
            matter power spectrum, and do not include 
			contributions from non-linear galaxy bias and connected parts of the covariance, hence the theory errors are 
			underestimated at small scales.
			The difference between the solid and dashed theory lines are due to the edge effects as estimated by $\mathcal 
			A_W$.
		\emph{Right panel}: Comparison between the different error estimation methods in mocks.
			$\mean{\text{Subsample}}$ refers to the mean and standard deviation across subsamples.
			$\mean{\text{Subsample-2}}$ is similar to $\mean{\text{Subsample}}$, except for each subsample we also count 
			the cross terms with other subsamples. This reduces edge effects but also 
			leads to correlations between different subsamples at large scales.
	  		}
	      \label{fig:qpm_wgg_err}
    	\end{figure*}
	
		In Fig.~\ref{fig:qpm_wgg_err} we compare the clustering error estimates using different estimators. The LS estimator yields 
		the lowest errors followed by the estimator in Eq.~\eqref{xi_GG}. The relative trends 
		between the estimators are consistent with the theory estimates from  expressions in 
		Eq.~\eqref{eq:general_cov_full} and Eq.~\eqref{eq:numerical_clustering_cov_full}
        and, more generally, with the idea that each time you substitute a zero-mean field with a
        field that has a non-zero mean, the variance increases. We caution that for the clustering measurements,
        we have not completely explored the 
        consistency between theory and empirical error estimates. Our theory estimates do not capture the full
        contributions from non-linear bias, redshift space distortions, 
        and the connected part of the covariance. 
        Also in 
		Eq.~\eqref{eq:general_cov_full} $\widehat P_{ij}(\vec k)=\widehat P(\vec k)$ and some terms from 
		Eq.~\eqref{eq:general_cov_full} 
		will be removed in the case of the estimator in Eq.~\eqref{xi_GG}; see Eq.~\eqref{eq:w_cov_full}.
		
		In Fig.~\ref{fig:qpm_wgg_err} we also compare the error estimates from jackknife and standard deviation across 
		different mock realizations, using different estimators. 
        Our results suggest that the jackknife 
		overestimates the errors at all scales, even when the scales are larger than the jackknife
        region size. This is contrary to the expectations, since at scales larger than the jackknife region size, the
        assumption that the regions are independent is violated and thus the errors are expected to be
        underestimated. However, since the jackknife 
		regions are much smaller than the survey size, the contribution from super-sample variance and other window 
		function-dependent terms is expected to be larger in the jackknife. 
		In the case of the LS estimator, our theory estimates (the difference between solid and dashed green lines) 
		suggest that the increase in error from edge effects is also
		important at large scale and can lead to the jackknife errors being over-estimated by $10-20\%$.
		For the non-optimal estimators, the 
		increased contribution from the additional terms identified in Appendix~\ref{appendix:covariance} dominates and 
		hence the increase in jackknife errors compared to the standard deviation is substantially more, when compared 
		to the LS estimator.
		           
\end{document}